\documentclass[acmsmall]{acmart}

\usepackage{color,soul}
\usepackage{proof}
\usepackage{amsmath}
\usepackage{amssymb}
\usepackage{algorithm}
\usepackage{algpseudocode}
\usepackage{amsthm}

\newcommand{\setof}[1]{\ensuremath{\left \{ #1 \right \}}}
\newcommand{\tuple}[1]{\ensuremath{\left \langle #1 \right \rangle }}
\newcommand{\before}[1]{\ensuremath{\bullet} #1}
\newcommand{\after}[1]{#1 \ensuremath{\bullet}}
\newcommand{\first}[1]{\ensuremath{\mathbf{first}(#1)}}
\newcommand{\pred}[1]{\ensuremath{\mathbf{pred}(#1)}}
\newcommand{\successors}[1]{\ensuremath{\mathbf{succ}(#1)}}
\newcommand{\traces}[1]{\ensuremath{\mathbf{traces}(#1)}}
\newcommand{\variables}[1]{\ensuremath{\mathbf{vars}(#1)}}
\newcommand{\subexpressions}[1]{\ensuremath{\mathbf{subs}(#1)}}
\newcommand{\dom}[1]{\ensuremath{\mathbf{dom}\;#1}}
\let\comment=\relax
\newcommand{\comment}[1]{}
\setcopyright{none}

\usepackage{booktabs}   
\usepackage{subcaption} 

\begin{document}

	\title[Dataflow Analysis With Prophecy and History Variables]{Dataflow Analysis With Prophecy and History Variables}

\author{Martin Rinard}
\affiliation{
	\institution{MIT EECS and MIT CSAIL}            
}
\email{rinard@csail.mit.edu}          

	\author{Austin Gadient}
	\affiliation{
	\institution{Aarno Labs}            
	}
	\email{agadient@mit.edu}          

	\begin{abstract} 
Leveraging concepts from state machine refinement proofs, we use prophecy variables,
which predict information about the future program execution, to enable forward reasoning 
for backward dataflow analyses. Drawing prophecy and history variables (concepts from the
dynamic execution of the program) from the same lattice as the static program analysis results,
we require the analysis results to satisfy both the dataflow equations and the transition
relations in the operational semantics of underlying programming language. This approach
eliminates explicit abstraction and concretization functions and promotes a more direct
connection between the analysis and program executions, with the connection taking
the form of a bisimulation relation between concrete executions and an augmented operational
semantics over the analysis results. We present several
classical dataflow analyses with this approach (live variables, very busy expressions,
defined variables, and reaching definitions) along with proofs that highlight how this
approach can enable more streamlined reasoning. To the best of our knowledge, we are the
first to use prophecy variables for dataflow analysis. 
\end{abstract}

	\maketitle
	
	\thispagestyle{empty}
	\pagestyle{fancy}
\fancyfoot{}
\section{Introduction}

Dataflow analysis is a classic field. Originally developed to enable
compiler 
optimizations~\cite{appel2004modern,muchnick1997advanced,DragonBook,cooper2011engineering,kennedy2001optimizing,Kildall73,KamU77}, 
over the last several decades 
it has evolved to solve problems in a wide range of fields including,
for example, program verification~\cite{Monniaux05,CousotCFMMMR05, farzan2013inductive, namjoshi2018impact, fischer2005joining, BergerettiC85}, program understanding~\cite{chugh2008dataflow, farzan2010compositional, farzan2013duet, ramsey2010hoopl, KinlochM94}, 
and computer security~\cite{abate2020trace, rajani2020expressiveness, deng2017securing, deng2018securing, RussoS10}. 

Early in the history of the field the question of the relationship between
the analysis results and program executions arose. One answer to this
question developed as follows~\cite{DragonBook,Kildall73,KamU77,CousotC77,LernerMRC05,Aldrich2019Correctness}.
First, define an operational semantics in which the program executes
commands $c$ that read and write program states $\sigma$ to produce
a sequence of states $\sigma_0, \ldots, \sigma_i, \ldots$, with each
state storing the values of variables at a 
corresponding program point $l_i:c_i$ (i.e., the
program location $l_i$ before the command $c_i$ executes). 
Second, define a lattice $\tuple{S,\leq}$ of dataflow facts $s \in S$ along with
an abstraction function $\alpha$ (where $s = \alpha(\sigma)$) that maps each 
program state $\sigma$ to a corresponding dataflow fact $s$
and a concretization function $\gamma$ (where $\sigma \in \gamma(s)$)
that maps each dataflow fact $s$ to the set of program states 
$\sigma$ that it abstracts. Together $\alpha$ and $\gamma$ 
form a Galois connection~\cite{CousotC77}.

A sound dataflow analysis guarantees the property that for all program
states $\sigma$, $\alpha(\sigma) \leq s$, where $s$ is the result
that the analysis produces at the corresponding program point 
for $\sigma$ (this property essentially requires the analysis
to take all execution paths into account).
A natural way to prove an analysis sound is by 
forward reasoning, operating by induction on the length
of the program execution, with the induction step proved via
a case analysis on the last command to execute~\cite{LernerMRC05}.

There are several anomalies with this approach.  
First, many classic program analyses (for example, live variables 
and very busy expressions~\cite{DragonBook}) are backward
analyses that maintain information not about the past execution 
but about the future execution.  Forward reasoning is 
often ineffective for reasoning about these analyses or proving their
soundness. Standard presentations of dataflow analysis therefore typically
focus on forward analyses, with backward analyses introduced later
as a kind of dual of forward analyses~\cite{DragonBook,Aldrich2019Examples}.

Second, many classic dataflow analyses (such as, for example, reaching 
definitions or available expressions~\cite{DragonBook}) 
maintain information about the past execution of the
program that is not present in the program states $\sigma$
that the standard operational semantics maintains. For example,
the standard operational semantics for simple imperative
languages maintains only the current values of variables.
These semantics leave no record of which command produced the current value. Reaching
definitions extracts information about which commands produce
values read by subsequent variable uses. 
This fact makes it impossible to construct an
abstraction function $\alpha(\sigma)$ that
operates on the standard program state $\sigma$, 
which records only variable values --- the reaching
definition information is not available in $\sigma$. 
A solution to this problem is to develop an instrumented
semantics that maintains this past information explicitly
in $\sigma$ to enable the construction of the 
abstraction function~\cite{LernerMRC05,CousotC92}.

\subsection{History and Prophecy Variables} 

The relationship between a concrete and abstract perspective on 
a computation has also been explored in the context of using
refinement mappings to prove forward simulation relations for 
verifying the correctness of a (concrete) implementation with respect to an (abstract) 
specification~\cite{AbadiL91}. In this context the specification
and implementation are both modeled as state machines, with 
simulation proofs (proving that each implementation action correctly simulates
some specification action) verifying that the implementation correctly
implements the specification. 

Stating the appropriate correctness conditions that the specification and
implementation must preserve often involves reasoning
about the past execution of the specification and/or implementation. 
To enable this reasoning, the formal framework uses {\em history variables}, 
i.e., additional state components that do not affect the externally visible
behavior of the state machine but rather simply record information about the 
past execution that can then be used to state and prove relevant correctness conditions. 
History variables were initially developed in the context of
program verification~\cite{OwickiG76} and have since been widely
used under a variety of names (e.g., auxiliary variables, ghost variables) in a range
of communities including the programming languages and program verification 
communities~\cite{OwickiG76,ZeeKR08, jung2018iris, de2019spy}.

In some situations, the specification and implementation make (typically
nondeterministic) choices at different points in their execution, with, for 
example, a natural specification making the choice before the implementation. 
In these situations it is often not possible to prove that the implementation
correctly implements the specification using the standard history variable and 
forward simulation proof mechanisms~\cite{AbadiL91}. One solution to this
problem is to introduce {\em prophecy variables}, which make predictions about
the future executions of state machines (typically the specification) to 
enable the correctness properties to be stated and proven~\cite{AbadiL91,LynchV95}.

\subsection{History and Prophecy Variables for Dataflow Analysis}

Inspired by the use of prophecy and history variables for proving
simulation relations as well as the unsatisfying treatment of 
backward and forward analyses in the standard 
dataflow analysis framework, we use prophecy and history variables
to formalize a new treatment of both backward and forward
dataflow analyses. 

Backward analyses augment the standard operational semantics of the underlying
programming language with prophecy variables that 
(typically nondeterministically) predict any information 
about the future execution of the program required to establish the
correspondence between the analysis and the execution. Because some of these
predictions may be incorrect, the analysis also augments the semantics with
{\em prophecy variable preconditions} that check prediction correctness
to filter out any executions with incorrect predictions. 
Forward analyses augment the standard operational semantics of the underlying 
programming language with history variables that record any information
about the past execution required to establish the correspondence between
the analysis and the execution.

With this formulation, the standard semantics operates on 
states $\tuple{l,\sigma}$ and the augmented semantics operates on augmented states
$\tuple{l,\sigma, \pi}$, where $l$ is the label of the next command
to execute. $\sigma$ records the standard state of the program (for example,
the values of the variables that the program manipulates), and 
$\pi$ is the prophecy or history variable from the analysis. The
standard operational semantics involves a transition relation
$\tuple{l,\sigma} \rightarrow \tuple{l',\sigma'}$; the augmented
operational semantics involves a transition relation
$\tuple{l,\sigma,\pi} \Rightarrow \tuple{l',\sigma',\pi'}$.
The introduction of the prophecy or history variable $\pi$ produces,
in effect, two executions of the program that run together in lockstep ---
the standard execution that operates on $\sigma$ and another 
execution that runs on top of the standard execution, 
may read both $\sigma$ and $\pi$, but only updates $\pi$. 

The dataflow analysis produces, for every program point $\before{l}$ (the
program point before the command at label $l$ executes) and $\after{l}$
(the program point after the command at label $l$ executes), analysis
results $\beta_{\before{l}}$ and $\beta_{\after{l}}$. These analysis
results are drawn from the same lattice as the prophecy or history
variables $\pi$, making it possible to substitute the analysis
results directly into the augmented operational semantics to obtain
transitions $\tuple{l,\sigma,\beta_{\before{l}}} \Rightarrow \tuple{l',\sigma',\beta_{\before{l'}}}$,
where $l'$ is the label of the command that executes next after the command at $l$. 

This setup enables us to formulate the soundness criteria that the dataflow analysis must preserve
via two properties that establish the correspondence between the dataflow
analysis and the program execution:
\begin{itemize}
\item {\bf Preservation}: $\tuple{l,\sigma,\pi}\Rightarrow\tuple{l',\sigma',\pi'}$
implies $\tuple{l,\sigma}\rightarrow\tuple{l',\sigma'}$. Preservation
ensures that the augmented semantics does not produce any new executions. 

\item {\bf Progress:} $\tuple{l,\sigma}\rightarrow\tuple{l',\sigma'}$ implies
$\tuple{l,\sigma, \beta_{\before{l}}}\Rightarrow\tuple{l',\sigma',\beta_{\before{l'}}}$. 
Progress requires prophecy variables to correctly predict
all possible future executions. In particular, proving $\tuple{l,\sigma, \beta_{\before{l}}}\Rightarrow\tuple{l',\sigma',\beta_{\before{l'}}}$
requires the analysis to produce analysis results $\beta_{\before{l}}$ that 
satisfy the prophecy variable preconditions that filter out
incorrect prophecy variable predictions.
For analyses that use history variables, Progress requires the 
history variables to correctly summarize all past executions. 

\end{itemize}

To satisfy these properties, the analysis must produce analysis results $\beta$ that satisfy
{\em both} the dataflow equations and the transition relations in the augmented
operational semantics. The analysis results $\beta$ therefore tie the analysis and concrete
executions together via the prophecy and history variables, with prophecy and history 
variable properties formalizing a direct connection between the analysis, the augmented semantics, 
and the standard semantics. This connection is reflected in the fact that, 
together, Preservation and Progress induce a bisimulation relation~\cite{Milner89,Benthem96}
between the standard semantics and the augmented semantics running on the 
analysis results $\beta_{\before{l}}$. This bisimulation relation formalizes
the guaranteed correspondence between the analysis and the standard execution
of the program. 

This connection can then be used to prove, via forward reasoning for both backward and
forward analyses, additional analysis properties. These analysis properties
may, for example, enable semantics-preserving program transformations
such as dead variable elimination (Section~\ref{sec:lv}), 
code hoisting (Section~\ref{sec:vbe}), or constant propagation (Section~\ref{sec:rd}), 
or to check for program correctness
properties such as the absence or presence of undefined variables (Section~\ref{sec:dv}).

\subsection{Contributions}

This paper makes the following contributions:
\begin{itemize}
\item {\bf Prophecy Variables for Backward Dataflow Analysis:} Prophecy variables were originally developed to prove forward 
simulation relations between state machines that take corresponding actions at different times.
Leveraging the ability of prophecy variables to predict information about the future execution,
we use prophecy variables to develop a unified treatment of backward and forward dataflow analyses.
In this treatment, backward analyses deliver accurate information about the future execution,
with prophecy variables enabling the statement and proof of precise conditions that the analysis must
satisfy to 1) accurately predict information about the future (as checked by the prophecy
variable preconditions) while 2) remaining consistent with the prophecy variable predictions. 
To the best of our knowledge we are the first to use prophecy variables for dataflow analysis. 

\item {\bf Mechanisms:} Drawing the prophecy and history variables from the same lattice
as the analysis results eliminates explicit abstraction and concretization functions
from the treatment, including the elimination of abstraction and concretization functions
from any proofs involving the analysis or the analysis results. The proofs instead
work with analysis-specific properties over the prophecy and history variables as induced
by the prophecy variable preconditions, prophecy variable predictions, and history
variable updates. These properties directly relate concrete executions and the analysis via a bisimulation
induced by the analysis results (Theorems~\ref{thm:lvpreservation}, \ref{thm:lvprogress}, 
\ref{thm:vbepreservation}, \ref{thm:vbeprogress}, \ref{thm:dvpreservation}, 
\ref{thm:dvprogress}, \ref{thm:rdpreservation}, and \ref{thm:rdprogress}). 

Replacing traditional collecting and/or instrumented semantics with explicit prophecy
or history variables leaves the standard operational semantics intact, separated from 
the prophecy and history variables in the augmented operational semantics. The result
is a more direct connection between the analysis and the concrete execution and the
elimination of the need to work through the instrumented and/or collecting semantics
to state and prove properties of standard program executions. 

\item {\bf Dataflow Analyses and Proofs:} We present several classical dataflow analyses
(live variables, Section~\ref{sec:lv}; very busy expressions, 
Section~\ref{sec:vbe}; defined variables, Section~\ref{sec:dv}; and reaching definitions,
Section~\ref{sec:rd}) with prophecy variables and history variables along with proofs that establish
the relevant bisimulations and proofs of analysis correctness properties for 
semantics-preserving program transformations. These proofs highlight the features
of our treatment, including the ability of prophecy variables to deliver a more unified
treatment of backward and forward dataflow analyses to enable forward reasoning for both backward and forward analyses.
They also highlight how the use of the same lattice for the prophecy variables, history variables, and
analysis results enables more streamlined reasoning. 

\end{itemize}

The remainder of the paper is structured as follows. Section~\ref{sec:overview} presents
an overview of the basic concepts in our treatment, including the Preservation and
Progress properties that together establish the bisimulation. Section~\ref{sec:lamguage}
presents the core imperative language that we use to present the dataflow analyses. 
Section~\ref{sec:backward} presents two backward analyses, live variables and very
busy expressions, including prophecy variable preconditions, prophecy variable 
predictions, proofs that establish the bisimulation between the analysis
results and program executions, and proofs that establish relevant analysis
correctness properties. Section~\ref{sec:forward} similarly presents two
forward dataflow analyses, defined variables and reaching definitions. 
We discuss related work in Section~\ref{sec:related} and conclude in 
Section~\ref{sec:conclusion}.

	\section{Overview}
\label{sec:overview}

We work with programs $P$ that contain labeled commands of the 
form $l : c \in P$, where $l,g \in L$, $c \in C$. 
$\textbf{labels}(P) = \{ l . l : c \in P \}$ is the set of
labels in $P$. Labels are unique --- no two labeled commands
in $P$ have the same label $l$.  An executing program operates on states $\sigma \in \Sigma$. 
The standard operational semantics is modeled by a 
program execution transition relation $\tuple{l,\sigma} \rightarrow \tuple{l',\sigma'}$. Execution 
starts at $\tuple{l_0, \sigma_0}$, where $l_0 = \mathbf{first}(P)$ is the label of
the first command to execute and $\sigma_0$ is the initial state. Execution 
terminates if it encounters an $l : \mathbf{halt}$ command. 

The standard operational semantics is typically defined by a set of program execution rules. 
Each rule starts with a standard program configuration $\tuple{l,\sigma}$ to
produce a next configuration $\tuple{l,\sigma} \rightarrow \tuple{l',\sigma'}$. Each 
rule has a set of preconditions that must be satisfied for the rule to execute.
If the execution encounters an error, at least one of the relevant preconditions will 
not be satisfied and the execution will become stuck in a configuration 
$\tuple{l,\sigma}$ such that $\tuple{l,\sigma} \not\rightarrow$. 

Each program analysis augments the state with a prophecy or history 
variable $\pi \in \Pi$, where $\tuple{\Pi, \leq}$ is a lattice
ordered by $\leq$ 
with least upper bound $\lor$ and greatest lower bound $\land$. 
The analysis also defines an augmented operational semantics by updating the
program execution rules to define an augmented program execution relation
$\tuple{l,\sigma,\pi} \Rightarrow \tuple{l',\sigma',\pi'}$. 

If $\pi$ is a prophecy variable, the updated 
program execution rules use $\pi$ to predict information about the
future program execution, with incorrect predictions filtered out by 
new prophecy variable preconditions that check that the prediction was correct
(with the execution becoming stuck if the prediction was not
correct). If $\pi$ is a history variable, the updated rules use $\pi$ to 
record information about the past program execution. 

We require the augmented program execution relation
$\tuple{l,\sigma,\pi} \Rightarrow \tuple{l', \sigma', \pi'}$
not to introduce new executions. More precisely,
we require the augmented program execution relation to satisfy the 
following preservation property:

\begin{definition}
\label{def:preservation}
(Preservation): If $\tuple{l,\sigma,\pi}\Rightarrow\tuple{l',\sigma',\pi'}$, then
$\tuple{l,\sigma}\rightarrow\tuple{l',\sigma'}$.
\end{definition}

Verifying the preservation property is typically straightforward as the 
updated program execution rules in the augmented operational semantics
typically have the same preconditions over $l$ and $\sigma$ and generate
the same $l'$ and $\sigma'$ as the corresponding program execution rules
in the standard operational semantics. 

For each labeled command $l:c \in P$, there are two program points: $\before{l}$ (the program point
before $l:c \in P$ executes) and $\after{l}$ (the program point after $l:c \in P$ executes). 
The program analysis produces an analysis result $\beta \in \Pi$ 
(drawn from same lattice $\tuple{\Pi,\leq}$ as the prophecy and history variables $\pi$)
at each program point.  Given a labeled command $l:c \in P$, $\beta_{\before{l}}$ is the program analysis 
result at the program point before $l:c\in P$ executes; 
$\beta_{\after{l}}$ is the program analysis 
result at the program point after $l:c\in P$ executes. 

Conceptually, the analysis is consistent with the augmented operational semantics
if it produces an analysis result that enables a corresponding transition in the
augmented operational semantics for each transition in the standard operational
semantics. We formalize this requirement with the following Progress property:

\begin{definition}
\label{def:progress}
(Progress): If $\tuple{l,\sigma}\rightarrow\tuple{l',\sigma'}$, then
$\tuple{l,\sigma, \beta_{\before{l}}}\Rightarrow\tuple{l',\sigma',\beta_{\before{l'}}}$.
\end{definition}

For analyses with prophecy variables, the progress
property requires the analysis to produce correct predictions 
about all possible future executions (in the sense that the analysis results satisfy the 
prophecy variable preconditions that check for incorrect predictions). 
For analyses with history variables, the progress
property requires the analysis to produce results that take all 
possible past executions into account. 
It is the responsibility of the analysis developer to ensure that the
preservation and progress properties hold, typically by proving 
corresponding preservation and progress theorems for the analysis.

The Progress property is typically verified by local reasoning, usually by a case analysis
on the command that generated the $\tuple{l,\sigma} \rightarrow \tuple{l',\sigma'}$
transition. For backward program analyses the Progress property 
can flip the direction of causality 
to enable forward reasoning --- reasoning from a chosen point in the computation forward
along the potential program execution paths to verify a relationship between the analysis and
the execution of the program. Examples of this forward reasoning for backwards analyses
that use prophecy variables include Theorems~\ref{thm:lvcharacterize}, \ref{thm:vbeca}, \ref{thm:vbech}.
Because forward reasoning is typically 
straightforward for forward analyses, the Progress property can effectively
unify reasoning approaches for forward and backward analyses. 

We note that if Preservation (Definition~\ref{def:preservation}) and Progress
(Definition~\ref{def:progress}) both hold, then the relation $\sim$ defined by
$\tuple{l,\sigma} \sim \tuple{l, \sigma, \beta_{\before{l}}}$ is a 
bisimulation relation~\cite{Milner89,Benthem96} between standard and augmented program configurations. 

For analyses with prophecy variables $\pi$, the following downward closure metarule is often
helpful in ensuring the progress property holds:
\begin{definition}(Downward Closure Metarule):
\label{def:dcm}
\[
\infer{\tuple{l, \sigma, \pi} \Rightarrow \tuple{l', \sigma', \pi''}}
      {\tuple{l, \sigma, \pi} \Rightarrow \tuple{l', \sigma', \pi'} \;\;\; \pi'' \leq \pi'}
\]
\end{definition}
Conceptually, moving down the lattice from $\pi'$ to $\pi''$ takes fewer program
executions into account, which happens 1) when the prophecy variable makes a prediction
about a future execution and 2) at program control flow split points for backward program
analyses (which typically use prophecy variables). 

For analyses with history variables, the following upward closure metarule is often
helpful in ensuring the progress property holds:
\begin{definition}(Upward Closure Metarule):
\label{def:ucm}
\[
\infer{\tuple{l, \sigma, \pi} \Rightarrow \tuple{l', \sigma', \pi''}}
      {\tuple{l, \sigma, \pi} \Rightarrow \tuple{l', \sigma', \pi'} \;\;\; \pi' \leq \pi''}
\]
\end{definition}
Conceptually, moving up the lattice from $\pi'$ to $\pi''$ takes more program
executions into account, which happens at program control flow join points for forward
program analyses (which typically use history variables).

	\section{Core Programming Language}
\label{sec:lamguage}

We present program analyses for a core programming language inspired by Glynn Winskell's imperative
language \textbf{IMP}~\cite{Winskel93}.  Notable differences include the 
introduction of labels for all commands and the use of variables $V$ instead of locations ($\mathbf{Loc}$). 

$n,m \in N$ is the set of integers. $t \in T = \setof{\mathbf{true},\mathbf{false}}$ is the set of truth values.
Programs work with variables $v,w \in V$, arithmetic expressions $e \in E$, and 
boolean expressions $b \in B$ defined as follows:
\[
\begin{array}{rcl}
E & ::= & n|v|E_0+E_1|E_0-E_1|E_0\times E_1 \\
B & ::= & \mathbf{true}|\mathbf{false}|E_0=E_1|E_0\leq E_1| \mathbf{not} \; B|B_0 \;\mathbf{and}\;B_1|B_0\;\mathbf{or}\;B_1 \\
C & ::= & \mathbf{skip}|v:=E|\mathbf{if}\;B\;\mathbf{then}\;g|\mathbf{goto}\; g|\mathbf{halt}|\mathbf{done}
\end{array}
\]
\noindent $\variables{e}$ is the set of variables $v$ that $e$ reads, 
$\variables{b}$ is the set of variables $v$ that $b$ reads, and
$\variables{c}$ is the set of variables that $c$ reads. 

Each program $P$ is a sequence of labeled commands of the 
form $l : c$, where $l,g \in L$, $c \in C$. 
Given a program $P$ and label $l \in \mathbf{labels}(P)$, $l' = \mathbf{next}(l)$ 
is the label $l'$ of the next command (in the sequential execution order) in $P$ after $l:c \in P$. 
Conceptually, when the program executes a $l:\mathbf{halt} \in P$ command, the program 
stops executing in the $\mathbf{done}$ state. We therefore require 
if $l : \mathbf{halt} \in P$ and $l' = \mathbf{next}(l)$, then $l':\mathbf{done} \in P$. 
We also require $\mathbf{next}(l) = g$ if $l:c = \mbox{if } l:\mathbf{goto}\;g \in P$
(but typically reference the branch target $g$ explicitly instead of $\mathbf{next}(l)$). 
For each labeled command $l:c \in P$, there are two program points: $\before{l}$ (the program point
before $l:c \in P$ executes) and $\after{l}$ (the program point after $l:c$ executes). 
We define the successors $\successors{l}$ of $l$ as follows:
\[
\successors{l} 
=
\begin{cases}
\setof{\mathbf{next}(l)} & \mbox{if }  
l:v:=e \in P, l:\mathbf{skip} \in P, \mbox{ or } l:\mathbf{halt} \in P\\
\setof{g} & \mbox{if }  l:\mathbf{goto}\; g \in P \\
\setof{\mathbf{next}(l), g} & \mbox{if } l:\mathbf{if}\;b\;\mathbf{then}\;g \in P \\
\end{cases}
\]
\noindent and the predecessors $\pred{l}$ of $l$ as $\pred{l} = \setof{g . \successors{g} = l}$.

\subsection{Standard Operational Semantics} 
\label{sec:standardOperationalSemantics}

A program in execution maintains a state $\sigma : V \rightarrow N \in \Sigma$, where $\sigma(v)$ is the
value of the variable $v$ in state $\sigma$. Given an arithmetic expression $e$ and state $\sigma$, 
we define the arithmetic expression evaluation relation $\tuple{e, \sigma} \rightarrow n$ as the smallest
relation (under subset inclusion) over $E \times \Sigma \times N$ that satisfies the arithmetic 
expression evaluation rules in Figure~\ref{fig:eeer}. Given a boolean expression $b$ and state $\sigma$,
we define the boolean expression evaluation relation $\tuple{b, \sigma} \rightarrow t$ as the
smallest relation (under subset inclusion) over $B \times \Sigma \times T$ that satisfies the
boolean expression evaluation rules in Figure~\ref{fig:beer}.

\begin{figure*}[t]
\begin{center}
\begin{tabular}{ccc}
$ \tuple{n,\sigma} \rightarrow n$
& 
$\infer{\tuple{v,\sigma} \rightarrow \sigma(v)}
               {v \in \dom{\sigma}}$
&
$\infer{\tuple{e_0+e_1,\sigma} \rightarrow n_0+n_1}
       {\tuple{e_0,\sigma}\rightarrow n_0\; \tuple{e_1,\sigma} \rightarrow n_1}$
\end{tabular}
\end{center}

\begin{center}
\begin{tabular}{cc}
$\infer{\tuple{e_0-e_1,\sigma} \rightarrow n_0-n_1}
       {\tuple{e_0,\sigma}\rightarrow n_0\; \tuple{e_1,\sigma} \rightarrow n_1}$
&
$\infer{\tuple{e_0*e_1,\sigma} \rightarrow n_0*n_1}
       {\tuple{e_0,\sigma}\rightarrow n_0\; \tuple{e_1,\sigma} \rightarrow n_1}$
\end{tabular}
\end{center}
\caption{\label{fig:eeer} Standard Arithmetic Expression Evaluation Rules}

	\begin{center}
		\begin{tabular}{ccc}
			$ \tuple{\mathbf{true},\sigma} \rightarrow \mathbf{true}$
			& 
			$ \tuple{\mathbf{false},\sigma} \rightarrow \mathbf{false} $
			&
			$\infer{\tuple{e_0=e_1,\sigma} \rightarrow n_0=n_1}
			{\tuple{e_0,\sigma}\rightarrow n_0\; \tuple{e_1,\sigma} \rightarrow n_1}$
		\end{tabular}
	\end{center}
	
	\begin{center}
		\begin{tabular}{ccc}
			$\infer{\tuple{e_0 \leq e_1,\sigma} \rightarrow n_0 \leq n_1}
			{\tuple{e_0,\sigma}\rightarrow n_0\; \tuple{e_1,\sigma} \rightarrow n_1}$
			&
			$ \infer{\tuple{\mathbf{not}\; b,\sigma} \rightarrow \mathbf{false}}
			{\tuple{b, \sigma}\rightarrow \mathbf{true}} $
			&
			$ \infer{\tuple{\mathbf{not}\; b,\sigma} \rightarrow \mathbf{true}}
			{\tuple{b, \sigma}\rightarrow \mathbf{false}} $
		\end{tabular}
	\end{center}

  \begin{center}
	\begin{tabular}{cc}
		$\infer{\tuple{b_0 \;\mathbf{and}\; b_1,\sigma} \rightarrow t_0 \;\mathbf{and}\; t_1}
		{\tuple{b_0,\sigma}\rightarrow t_0\; \tuple{b_1,\sigma} \rightarrow t_1}$
		&
		$\infer{\tuple{b_0 \;\mathbf{or}\; b_1,\sigma} \rightarrow t_0 \;\mathbf{or}\; t_1}
         {\tuple{b_0,\sigma}\rightarrow t_0\; \tuple{b_1,\sigma} \rightarrow t_1}$
    \end{tabular}
  \end{center}
	\caption{\label{fig:beer} Standard Boolean Expression Evaluation Rules}

  \begin{center}
	\begin{tabular}{c}
		$\infer{\tuple{l,\sigma}\rightarrow \tuple{\mathbf{next}(l) ,\sigma[v\mapsto n]}}
		{l:v:=e\;\in\; \mathbf{P}\;\;\tuple{a,\sigma}\rightarrow n}$
	\end{tabular}
  \end{center}

  \begin{center}
	\begin{tabular}{cc}
		$\infer{\tuple{l,\sigma}\rightarrow \tuple{\mathbf{next}(l), \sigma}}
		{l\;:\;\mathbf{if}\;b\;\mathbf{then}\;g\;\in\; \mathbf{P}\;\;\; \tuple{b,\sigma} \rightarrow \mathbf{false}}$
		&
		$\infer{\tuple{l,\sigma}\rightarrow \tuple{g, \sigma}}
		{l:\mathbf{if}\;b\;\mathbf{then}\;g\;\in\; \mathbf{P}\;\;\; \tuple{b,\sigma} \rightarrow \mathbf{true}}$
	\end{tabular}
  \end{center}

  \begin{center}
	\begin{tabular}{ccc}
		$\infer{\tuple{l,\sigma} \rightarrow \tuple{g, \sigma}}
		{l:\mathbf{goto}\;g\;\in\; \mathbf{P}}$ 
                &
		$\infer{\tuple{l,\sigma} \rightarrow \tuple{\mathbf{next}(l),\sigma}}
		{l:\mathbf{skip}\;\in\; \mathbf{P}}$
		&
		$\infer{\tuple{l,\sigma} \rightarrow \tuple{\mathbf{next}(l),\sigma}}
                {l : \mathbf{halt} \in P \;\;\; \mathbf{next}(l):\mathbf{done} \in P}$
	\end{tabular}
  \end{center}

\caption{\label{fig:per} Standard Program Execution Rules}
\end{figure*}

\begin{figure*}[t]
\begin{center}
\begin{tabular}{ccc}
$ \tuple{n,\sigma, \pi} \Rightarrow n$
& 
$\infer{\tuple{v,\sigma, \pi} \Rightarrow \sigma(v)}
               {v \in \dom{\sigma}}$
&
$\infer{\tuple{a_0+a_1,\sigma, \pi} \Rightarrow n_0+n_1}
       {\tuple{a_0,\sigma, \pi}\Rightarrow n_0\; \tuple{a_1,\sigma, \pi} \Rightarrow n_1}$
\end{tabular}
\end{center}

\begin{center}
\begin{tabular}{cc}
$\infer{\tuple{a_0-a_1,\sigma, \pi} \Rightarrow n_0-n_1}
       {\tuple{a_0,\sigma, \pi}\Rightarrow n_0\; \tuple{a_1,\sigma, \pi} \Rightarrow n_1}$
&
$\infer{\tuple{a_0*a_1,\sigma, \pi} \Rightarrow n_0*n_1}
       {\tuple{a_0,\sigma, \pi}\Rightarrow n_0\; \tuple{a_1,\sigma, \pi} \Rightarrow n_1}$
\end{tabular}
\end{center}
\caption{\label{fig:aeeer} Baseline Augmented Arithmetic Expression Evaluation Rules}

	\begin{center}
		\begin{tabular}{ccc}
			$ \tuple{\mathbf{true},\sigma, \pi} \Rightarrow \mathbf{true}$
			& 
			$ \tuple{\mathbf{false},\sigma, \pi} \Rightarrow \mathbf{false} $
			&
			$\infer{\tuple{a_0=a_1,\sigma, \pi} \Rightarrow n_0=n_1}
			{\tuple{a_0,\sigma, \pi}\Rightarrow n_0\; \tuple{a_1,\sigma, \pi} \Rightarrow n_1}$
		\end{tabular}
	\end{center}
	
	\begin{center}
		\begin{tabular}{ccc}
			$\infer{\tuple{a_0 \leq a_1,\sigma, \pi} \Rightarrow n_0 \leq n_1}
			{\tuple{a_0,\sigma, \pi}\Rightarrow n_0\; \tuple{a_1,\sigma, \pi} \Rightarrow n_1}$
			&
			$ \infer{\tuple{\mathbf{not}\; b,\sigma, \pi} \Rightarrow \mathbf{false}}
			{\tuple{b, \sigma, \pi}\Rightarrow \mathbf{true}} $
			&
			$ \infer{\tuple{\mathbf{not}\; b,\sigma, \pi} \Rightarrow \mathbf{true}}
			{\tuple{b, \sigma, \pi}\Rightarrow \mathbf{false}} $
		\end{tabular}
	\end{center}

  \begin{center}
	\begin{tabular}{cc}
		$\infer{\tuple{b_0 \;\mathbf{and}\; b_1,\sigma, \pi} \Rightarrow t_0 \;\mathbf{and}\; t_1}
		{\tuple{b_0,\sigma, \pi}\Rightarrow t_0\; \tuple{b_1,\sigma, \pi} \Rightarrow t_1}$
		&
		$\infer{\tuple{b_0 \;\mathbf{or}\; b_1,\sigma, \pi} \Rightarrow t_0 \;\mathbf{or}\; t_1}
         {\tuple{b_0,\sigma, \pi}\Rightarrow t_0\; \tuple{b_1,\sigma, \pi} \Rightarrow t_1}$
    \end{tabular}
  \end{center}
	\caption{\label{fig:abeerrd} Baseline Augmented Boolean Expression Evaluation Rules}

  \begin{center}
	\begin{tabular}{cc}
		$\infer{\tuple{l,\sigma,\pi}\Rightarrow \tuple{\mathbf{next}(l), \sigma,\pi}}
		{l\;:\;\mathbf{if}\;b\;\mathbf{then}\;g\;\in\; \mathbf{P}\;\;\; \tuple{b,\sigma,\pi} \Rightarrow \mathbf{false}}$
		&
		$\infer{\tuple{l,\sigma,\pi}\Rightarrow \tuple{g, \sigma,\pi}}
		{l:\mathbf{if}\;b\;\mathbf{then}\;g\;\in\; \mathbf{P}\;\;\; \tuple{b,\sigma,\pi} \Rightarrow \mathbf{true}}$
	\end{tabular}
  \end{center}

  \begin{center}
	\begin{tabular}{c}
		$\infer{\tuple{l,\sigma,\pi}\Rightarrow \tuple{\mathbf{next}(l) ,\sigma[v\mapsto n], \pi[v\mapsto \setof{l}]}}
		{l:v:=e\;\in\; \mathbf{P}\;\;\tuple{a,\sigma,\pi}\Rightarrow n}$
	\end{tabular}
  \end{center}

  \begin{center}
	\begin{tabular}{ccc}
		$\infer{\tuple{l,\sigma,\pi} \Rightarrow \tuple{g, \sigma,\pi}}
		{l:\mathbf{goto}\;g\;\in\; \mathbf{P}}$
		&
		$\infer{\tuple{l,\sigma,\pi} \Rightarrow \tuple{\mathbf{next}(l),\sigma,\pi}}
		{l:\mathbf{skip}\;\in\; \mathbf{P}}$
		&
		$\infer{\tuple{l,\sigma,\pi} \Rightarrow \tuple{\mathbf{next}(l),\sigma,\pi}}
                {l : \mathbf{halt} \in P \;\;\; \mathbf{next}(l):\mathbf{done} \in P}$
	\end{tabular}
  \end{center}

\caption{\label{fig:aper} Baseline Augmented Program Execution Rules}
\end{figure*}

Given a program $P$, the standard operational semantics works with configurations of the form $\tuple{l, \sigma}$, 
where $l$ is the label of a labeled command $l:c \in P$ and $\sigma : V \rightarrow N$ is an environment
that maps variables $v \in V$ to values $n \in N$. We define 
the program execution relation $\tuple{l, \sigma} \rightarrow \tuple{l', \sigma'}$ 
as the smallest relation (under subset inclusion) over
$L \times \Sigma \times L \times \Sigma$ that satisfies the program execution rules
in Figure~\ref{fig:per}.

Given a program $P$, a sequence of configurations $s = \tuple{l_0, \sigma_0} \rightarrow \cdots \rightarrow \tuple{l_n, \sigma_n} \cdots$,
where $l_0 = \first{P}$ and $\sigma_0 = \emptyset$ is the initial state of $P$, is an {\em execution} of $P$.
If $s$ is a finite sequence of length $n+1$, then $s$ is a {\em finite execution of $P$} and 
$\tuple{l_n, \sigma_n}$ is the last element in the sequence. 
If $l_n : \mathbf{done} \in P$, then the sequence is a {\em complete execution} of $P$, otherwise, it is a {\em partial execution} of $P$. 
If $s$ is a partial execution of $P$ and $\tuple{l_n, \sigma_n} \not\rightarrow$, then $s$ is a {\em stuck execution} of $P$
(an execution of $P$ can become stuck if the precondition of an expression evaluation or command execution rule is not satisfied).
If $s$ is an infinite sequence, then the sequence is an {\em infinite} execution of $P$. 
The $\traces{P, \rightarrow}$ are all of the complete and infinite executions of $P$.

\subsection{Baseline Augmented Operational Semantics}
\label{sec:baos}

Each program analysis typically updates only a few program execution rules,
with the remaining rules simply threading the prophecy or history variable $\pi$
through the execution unchanged. We therefore define an baseline augmented
operational semantics by updating all of the rules from the standard
operational semantics (Figures~\ref{fig:eeer}, \ref{fig:beer}, and \ref{fig:per})
to simply thread $\pi$ through the execution unchanged (by changing $\sigma$ to $\sigma, \pi$
in each rule) (only for completeness, presented in Figures~\ref{fig:aeeer}, \ref{fig:abeerrd}, and \ref{fig:aper}). 
Each analysis then updates one or more of the rules
from the baseline augmented operational semantics to appropriately update and/or check the
prophecy or history variable $\pi$ as appropriate for that analysis. 

Analyses that use history variables $\pi$ typically record
actions taken during the execution of the program. In this case augmented executions of 
$P$ never become stuck because of the augmentation. 
Analyses that use prophecy variables $\pi$, on the other hand, typically make nondeterministic 
predictions that are validated later in the execution. Executions involving
invalid predictions become stuck at the preconditions that validate the
predictions. 

\comment{
We note that, for any $\pi$, the baseline augmented operational semantics satisfies the 
Preservation (Definition~\ref{def:preservation}) and Progress (Definition~\ref{def:progress})
properties --- the rules are identical except for threading $\pi$ through the execution. 
}

	\section{Backward Analyses and Prophecy Variables}
\label{sec:backward}

We next present several backward analyses, including the use of prophecy variables
to formulate and prove properties characterizing the relationship between the
analysis and zprogram execution. 

\subsection{Live Variables}~\label{sec:lv}

Live variables analysis (conservatively) determines, for each program point,
the variables that are {\em live} at that program point, i.e., variables that may
be read before they are written in the future program execution. The analysis
uses a backward dataflow analysis to reason about the future execution of
the program. It therefore augments the standard operational semantics with a prophecy
variable $\pi \subseteq V$. $\pi$ predicts which variables are live; i.e., 
which variables will be read by some future command before the variable
is reassigned. The prophecy variable $\pi$ is drawn from the 
live variables program analysis lattice $\tuple{\Pi,\subseteq}$,
where $\Pi = \mathcal{P}(E)$ is ordered under subset 
inclusion ($\subseteq$), with least upper bound $\cup$ and greatest lower bound
$\cap$. 

\subsubsection{Augmented Operational Semantics} 
\label{sec:lvaos}
Starting with the baseline augmented operational semantics (Section~\ref{sec:baos}),
which passes the prophecy variable $\pi$ unchanged through all commands, 
the analysis updates the program execution rules that read variables  
(i.e., the rules for commands $l: v := e \in P$ and $l: \mathbf{if}\; b \; \mathbf{then} \; g \in P$)
to include new prophecy variable predictions that predict which variables will be live after the 
variable reads. 
Conceptually, the rule for $l: v := e \in P$ adds $v$ to the set of predicted live variables, 
then predicts that some subset of $\pi\cup\setof{v}$ will no longer be live after 
$l: v := e \in P$. Some variables may become dead either because 
$e$ contained the last read to a variable before the variable is reassigned
or because $v$ itself is not read before it is reassigned. 
Similarly, the rule for $l: \mathbf{if}\; b \; \mathbf{then} \; g \in P$ predicts that
some subset of the predicted live variables $\pi$ before 
$l: \mathbf{if}\; b \; \mathbf{then} \; g \in P$
will no longer be live after $l$,
for example because $b$ contained the last access to a variable before the
variable is reassigned. We do not apply the downward closure metarule (Definition~\ref{def:dcm}).

\[
\infer{\tuple{l,\sigma, \pi}\Rightarrow \tuple{\mathbf{next}(l), \sigma[v\mapsto n], \pi'}}
      {l:v:=e\;\in\; \mathbf{P}\;\;\tuple{a,\sigma, \pi}\Rightarrow n\;\;\; \pi' \subseteq \pi \cup \setof{v}}
\]
\[
\infer{\tuple{l,\sigma, \pi}\Rightarrow \tuple{\mathbf{next}(l), \sigma, \pi'}}
      {l:\mathbf{if}\;b\;\mathbf{then}\;g\;\in\; \mathbf{P}\;\;\; \tuple{b,\sigma, \pi} \Rightarrow \mathbf{false}\;\;\; \pi' \subseteq \pi}
\]

\[
\infer{\tuple{l,\sigma, \pi}\Rightarrow \tuple{g, \sigma, \pi'}}
      {l:\mathbf{if}\;b\;\mathbf{then}\;g\;\in\; \mathbf{P}\;\;\; \tuple{b,\sigma, \pi} \Rightarrow \mathbf{true} \;\;\; \pi' \subseteq \pi}
\]

Of course, it is possible for the program execution rules to 
mispredict which variables will be dead after executing
$l : v := e \in P$ or $l: \mathbf{if}\; b \; \mathbf{then} \; g \in P$. 
The augmented operational semantics therefore updates the variable read rule 
with the prophecy variable precondition $v \in \pi$, which requires
that every variable $v$ read during expression evaluation must be
predicted live. With this precondition, all executions that mispredict
a live variable become stuck at the command that attempts to read
the mispredicted variable. Here $\dom{\sigma}$ is the domain
of $\sigma$ viewed as a function --- the set of variables $v$ for 
which $\sigma(v)$ is defined. 

\[
\infer{\tuple{v,\sigma,\pi} \Rightarrow \sigma(v)}
      {v \in \dom{\sigma}\;\;\; v \in \pi}
\]

We next state some lemmas and the Preservation theorem. These lemmas and theorems
essentially leverage/formalize the fact that the introduction of the prophecy
variable $\pi$ into the augmented operational semantics (conceptually) 
preserves the standard operational semantics as long as the prophecy
variable preconditions $v \in \pi$ encountered during the evaluation
of an expression are satisfied:
\begin{lemma}
\label{lem:lvabee}
If $\tuple{e, \sigma, \pi} \Rightarrow n$, then $\tuple{e, \sigma} \rightarrow n$.
If $\tuple{b, \sigma, \pi} \Rightarrow t$, then $\tuple{b, \sigma} \rightarrow t$.
\end{lemma}
\noindent{\bf Proof:} 
The updated expression evaluation rules in the 
augmented operational semantics have the same preconditions 
(with the exception of the prophecy variable preconditions $v \in \pi$, which
are not present in the standard operational semantics) 
and produce the same expression values as the corresponding 
rules from the standard operational semantics. 
\noindent$\blacksquare$

\begin{lemma}
\label{lem:lvbaee}
If $\tuple{e, \sigma} \rightarrow n$ and $\variables{e} \subseteq \pi$, then $\tuple{e, \sigma, \pi} \Rightarrow n$.
If $\tuple{b, \sigma} \rightarrow t$ and $\variables{b} \subseteq \pi$, then $\tuple{b, \sigma, \pi} \Rightarrow t$.
\end{lemma}
\noindent{\bf Proof:} $\variables{e} \subseteq \pi$ and $\variables{b} \subseteq \pi$
ensure that any prophecy variable precondition $v \in \pi$
from the local variable read rule is satisfied during the evaluation of $e$ or $b$, 
so $\tuple{e, \sigma, \pi} \Rightarrow n'$ for some $n'$
and $\tuple{b, \sigma, \pi} \Rightarrow t'$ for some $t'$. 
By Lemma~\ref{lem:lvabee}, $n' = n$ and $t' = t$. 
\noindent$\blacksquare$

\begin{lemma}
\label{lem:lvens}
If $\tuple{l,\sigma,\pi} \Rightarrow \tuple{l',\sigma',\pi'}$ and $l:c \in P$, then $\variables{c} \subseteq \pi$.
\end{lemma}
\noindent{\bf Proof:} Consider any variable $v \in \variables{c}$. 
If $v$ is not in the set of predicted live variables $\pi$, the prophecy variable precondition $v \in \pi$
of the local variable read rule will not be satisfied during the evaluation of an
expression $e$ or $b$ in $c$, the evaluation of $e$ or $b$ will 
become stuck, the execution of $l:c \in P$ will become stuck, and $\tuple{l,\sigma,\pi} \not\Rightarrow$.
\noindent$\blacksquare$

\begin{lemma} 
\label{lem:lvvup}
If $\tuple{l,\sigma,\pi} \Rightarrow \tuple{l', \sigma', \pi'}$, 
$v \not\in \pi$, and $v \in \pi'$, then $l:v:=e \in P$: 
\end{lemma}
\noindent{\bf Proof:} 
Case analysis on $l:c\in P$:
\begin{itemize}
\item $l:c = l:\mathbf{if}\; b \; \mathbf{then}\; g \in P$: 
In this case $\pi' \subseteq \pi$, so there is no $v$ such that $v \not\in \pi$ and $v \in \pi'$.

\item $l:c = l:\mathbf{goto}\; g \in P$, $l:c = l:\mathbf{skip} \in P$ or $l:c = l:\mathbf{halt} \in P$:
In this case $\pi' = \pi$, so there is no $v$ such that $v \not\in \pi$ and $v \in \pi'$.

\item $l:c = l:v:=e \in P$: 
This is the only remaining case. In this case $\pi' \subseteq \pi \cup \setof{v}$ and it 
is possible for $v \not\in \pi$ and $v \in \pi'$.
\end{itemize}
\noindent$\blacksquare$

\begin{theorem} 
\label{thm:lvpreservation}
(Preservation)
If $\tuple{l, \sigma, \pi} \Rightarrow \tuple{l', \sigma', \pi'}$ then 
$\tuple{l, \sigma} \rightarrow \tuple{l', \sigma'}$. 
\end{theorem}
\noindent{\bf Proof:} All rules that generate a transition
$\tuple{l, \sigma, \pi} \Rightarrow \tuple{l', \sigma', \pi'}$
in the augmented operational semantics
have the same preconditions over $l:c \in P$ and $\sigma$ and 
produce the same values for $l'$ and $\sigma'$ as the
corresponding rules from the standard operational semantics. 
\noindent$\blacksquare$

\subsubsection{Live Variables Analysis}

The live variables analysis is a backward dataflow analysis that propagates variable liveness
information backward against the flow of control. 
For each command $l:c \in P$, the analysis produces $\beta_{\before{l}} \subseteq V$ (the set of
variables live at the program point before $l:c \in P$) and 
$\beta_{\after{l}} \subseteq V$ (the set of variables live after $l:c \in P$). 
The analysis obtains the $\beta_{\before{l}}$ and $\beta_{\after{l}}$ 
by formulating and solving, using standard least fixed-point
techniques, the following set of backward dataflow equations:
\[
\begin{array}{rcl}
\beta_{\after{l}} & = & \emptyset \mbox{ if } l:\mathbf{halt} \in P \\
\beta_{\after{l}} & = &  \cup \beta_{\before{g}}, \mbox{ where } l:c \in P \mbox{ and } g \in \successors{l} \\
\beta_{\before{l}} & = &  f(l, \beta_{\after{l}}) \\
\end{array}
\]
\noindent where $f$ is the transfer function for the analysis defined as follows:

\[
f(l, \beta) = 
\begin{cases}
(\beta - \setof{v}) \cup \variables{e} & \mbox{if }  l:v:=e \in P \\
\beta \cup \variables{b} & \mbox{if }  l:\; \mbox{if}\; b \; \mbox{then}\; l' \in P\\
\beta & \mbox{otherwise}
\end{cases}
\]

\noindent Note that these dataflow equations ensure that every variable $v$ read by a command $c$ is 
live when the command executes:
\begin{lemma}
\label{lem:lvvin}
$\variables{c} \subseteq \beta_{\before{l}}$ where $l:c \in P$.
\end{lemma}
\noindent{\bf Proof:} 
Case analysis on $l:c\in P$:
\begin{itemize}

\item $l:c = l:v:=e \in P$: \\
Then $\beta_{\before{l}} = (\beta_{\after{l}} - \setof{v}) \cup \variables{e}$ and $\variables{c} = \variables{e} \subseteq \beta_{\before{l}}$.

\item $l:c = l:\mathbf{if}\; b \; \mathbf{then}\; g \in P$: \\
Then $\beta_{\before{l}} = (\beta_{\after{l}} - \setof{v}) \cup \variables{b}$ and $\variables{c} = \variables{b} \subseteq \beta_{\before{l}}$.

\item $l:c = l:\mathbf{goto}\; g \in P$, $l:c = l:\mathbf{skip} \in P$, or $l:c = l:\mathbf{halt} \in P$: \\
Then $\variables{c} = \emptyset \subseteq \beta_{\before{l}}$.
\end{itemize}
\noindent$\blacksquare$

\subsubsection{Prophecy Variable Predictions and Dataflow Analyses with Sets of Elements and Subset Inclusion}
\label{sec:pvconsistentsubseteq}

Live variables is an instance of a more general class of backward dataflow analyses
in which the dataflow facts $\beta \in \Pi$ are sets of elements
with the dataflow lattice $\tuple{\Pi, \leq}$ ordered by subset inclusion ($\subseteq$),
with least upper bound $\cup$, greatest lower bound $\cap$, and
dataflow equations of the following form:
\[
\begin{array}{rcl}
\beta_{\after{l}} & = &  \cup \beta_{\before{g}}, \mbox{ where } l:c \in P \mbox{ and } g \in \successors{l} \\
\beta_{\before{l}} & = &  (\beta_{\after{l}} - D_l) \cup U_l \\
\end{array}
\]
\noindent where $D_l \in \Pi$ is the definition set for $l:c \in P$ and $U_l \in \Pi$ is the use set for $l:c \in P$.
Note that the transfer function for $l:c \in P$ is $f(l,\beta) = (\beta \mbox{ or }- D_l) \cup U_l$ and 
the equations ensure $\beta_{\before{l'}} \subseteq \beta_{\after{l}}$ for $l' \in \successors{l}$.
For live variables $D_l = \setof{v}$ and
$U_l = \variables{e}$ when $l:v:=e \in P$; $D_l = \emptyset$ and $U_l = \variables{b}$ when
$l:\; \mbox{if}\; b \; \mbox{then}\; l' \in P$. 
For $l:c = l:\mathbf{goto}\; g \in P$,
$l:c = l:\mathbf{skip} \in P$, and $l:c = l:\mathbf{halt} \in P$, 
$D_l = \emptyset$ and $U_l = \emptyset$.

One of the proof obligations required to show $\tuple{l,\sigma,\beta_{\before_{l}}} \Rightarrow \tuple{l', \sigma', \beta_{\before{l'}}}$ 
is establishing that the analysis results $\beta_{\before{l}}$ and $\beta_{\before{l'}}$ are consistent with
the prophecy variable predictions. We next show that prophecy variable predictions $\pi' \subseteq \pi \cup D_l$ are consistent with 
these analyses:
\begin{lemma}
\label{lem:pvconsistentsubseteq}
If $\beta_{\before{l}} = (\beta_{\after{l}} - D_l) \cup U_l$ and $\beta_{\before{l'}} \subseteq \beta_{\after{l}}$, 
then $\beta_{\before{l'}} \subseteq \beta_{\before{l}} \cup D_l$. 
\end{lemma}
\noindent{\bf Proof:} 
\begin{itemize}
\item  Known facts from dataflow analysis:
$\beta_{\before{l}} = (\beta_{\after{l}} - D_l) \cup U_l$ and $\beta_{\before{l'}} \subseteq \beta_{\after{l}}$.
\item Then $\beta_{\before{l}} \supseteq (\beta_{\before{l'}} - D_l) \cup U_l$, $\beta_{\before{l}} \supseteq (\beta_{\before{l'}} - D_l)$, 
$\beta_{\before{l}} \cup D_l \supseteq (\beta_{\before{l'}} - D_l) \cup D_l$,
$\beta_{\before{l}} \cup D_l \supseteq \beta_{\before{l'}} \cup D_l$, and
$\beta_{\before{l}} \cup D_l \supseteq \beta_{\before{l'}}$.
\end{itemize}
\noindent$\blacksquare$

If $D_l = U_l = \emptyset$ so that $f(l,\beta) = \beta$ and $\setof{l'} = \successors{l}$,
then $\beta_{\before{l}} = \beta_{\after{l}} = \beta_{\before{l'}}$. For live 
variables this is the case for $l:c = l:\mathbf{goto}\; g \in P$,
$l:c = l:\mathbf{skip} \in P$, and $l:c = l:\mathbf{halt} \in P$. 
In this case the analysis results are consistent with prophecy
variable predictions $\pi' = \pi$ and $\pi' \subseteq \pi$
(as in, for example, analyses that use the downward closure metarule):
\begin{lemma}
\label{lem:pvconsistenteq}
If $\beta_{\before{l}} = \beta_{\after{l}}$ and $\beta_{\after{l}} = \beta_{\before{l'}}$, then
$\beta_{\before{l}} = \beta_{\before{l'}}$ and $\beta_{\before{l'}} \subseteq \beta_{\before{l}}$. 
\end{lemma}
\noindent$\blacksquare$

\subsubsection{Live Variables Progress Theorem}

We next state and prove the Progress theorem for the live variables analysis. 
First, the proof must ensure that the analysis results $\beta_{\before{l}}$
satisfy the prophecy variable precondition $v \in \pi$ in the 
augmented expression evaluation rules for all 
variables $v$ in evaluated expressions $e$ and $b$. 
This property shows up as proof obligations of the
form $\variables{e} \subseteq \beta_{\before{l}}$ and 
$\variables{b} \subseteq \beta_{\before{l}}$ for
expressions $e$ and $b$ that appear in commands $l:c \in P$. 
These proof obligations are immediately discharged because 
the transfer function $f$ explicitly places $\variables{e}$ and
$\variables{b}$ in $\beta_{\before{l}}$ for commands $l:c \in P$
that contain $e$ or $b$ --- in other words, the 
prophecy variable precondition proof obligations are immediately discharged regardless of
the values of related program analysis results $\beta_{\after{l}}$ 
and $\beta_{\before{l'}}$ where $l' \in \successors{l}$. 

The proof must also ensure that the analysis results are consistent with
the prophecy variable predictions in the augmented operational semantics. 
For commands $l:c = l:v:=e\in P$ this property shows up as proof obligations 
$\beta_{\before{l'}} \subseteq \beta_{\before{l}} \cup \setof{v}$. For 
$:c = l:\mathbf{if}\; b \; \mathbf{then}\; g \in P$ this property shows up
as proof obligations $\beta_{\before{l'}} \subseteq \beta_{\before{l}}$. 
For $l:c = l:\mathbf{goto}\; g \in P$, $l:c = l:\mathbf{skip} \in P$, and
$l:c = l:\mathbf{halt} \in P$, this property shows up
as proof obligations $\beta_{\before{l'}} = \beta_{\before{l}}$. 
Unlike the prophecy variable precondition proof obligations, these
prophecy variable prediction proof obligations do depend on the relationship between 
$\beta_{\before{l}}$, $\beta_{\after{l}}$, and $\beta_{\before{l'}}$ where $l' \in \successors{l}$.
They can therefore be discharged by pushing the analysis result
$\beta_{\before{l'}}$ through the transfer function $f$ for $l:c \in P$ 
to check that the analysis related analysis results $\beta_{\before{l}}$,
$\beta_{\after{l}}$, and $\beta_{\before{l'}}$ are consistent with the prophecy variable
predictions. 

For live variables, the analysis and prophecy variable prediction
conform to the requirements of Lemmas~\ref{lem:pvconsistentsubseteq} and \ref{lem:pvconsistenteq}. 
So the prophecy variable prediction proof obligations for these commands
are immediately discharged by applying these lemmas. 

At a higher level, these properties ensure that the analysis never spontaneously takes a 
variable that is not live and makes it live. The augmented
operational semantics uses the prophecy variable $\pi$ to enforce this property, 
which must be preserved by the static analysis for the analysis to produce an analysis result
consistent with the prophecy variable predictions and in which the prophecy variable
preconditions hold. 

\begin{theorem}
\label{thm:lvprogress}
(Progress)
If $\tuple{l, \sigma} \rightarrow \tuple{l', \sigma'}$ then 
$\tuple{l, \sigma, \beta_{\before{l}}} \Rightarrow \tuple{l', \sigma', \beta_{\before{l'}}}$.
\end{theorem}
\noindent{\bf Proof:} 
If all of the prophecy variable preconditions are satisfied, 
the standard and augmented program execution rules for $l:c \in P$ define the
same values for $l'$, $\sigma'$, and all evaluated expressions $e$ or $b$. 
The following case analysis on $l:c \in P$ shows that the prophecy variable 
preconditions (which require all variables $v$ read in evaluated expressions
$e$ and $b$ to be predicted live) are satisfied and that 
$\beta_{\before{l}}$ and $\beta_{\before{l'}}$ satisfy 
$\tuple{l, \sigma, \beta_{\before{l}}} \Rightarrow \tuple{l', \sigma', \beta_{\before{l'}}}$:

\begin{itemize}
\item $l:c = l:v:=e \in P$:
\begin{itemize}
\item Facts from dataflow equations: \\
$\beta_{\before{l}} = (\beta_{\after{l}} - \setof{v}) \cup \variables{e}$ (from transfer function $f$ for $l:v:=e \in P$) and \\
$\beta_{\after{l}} = \beta_{\before{l'}}$ (because $\setof{l'} = \successors{l})$. 

\item By Lemma~\ref{lem:lvbaee}, $\tuple{l, \sigma} \rightarrow \tuple{l', \sigma'}$ and $\variables{e} \subseteq \beta_{\before{l}}$ 
imply $\tuple{e,\sigma,\beta_{\before{l}}} \Rightarrow n$, where $\tuple{e,\sigma} \rightarrow n$.

\item Prove prophecy variable precondition $\forall v \in \variables{e} . v \in \beta_{\before{l}}$, i.e., prove $\variables{e}  \subseteq \beta_{\before{l}}$: \\
$\variables{e} \subseteq (\beta_{\after{l}} - \setof{v}) \cup \variables{e} = \beta_{\before{l}}$. 

\item Prove consistent with prophecy variable prediction $\beta_{\before{l'}} \subseteq \beta_{\before{l}} \cup \setof{v}$: 
Lemma~\ref{lem:pvconsistentsubseteq}.
\comment{
Because $\beta_{\before{l}} = (\beta_{\after{l}} - \setof{v}) \cup \variables{e}$ and $\beta_{\after{l}} = \beta_{\before{l'}}$, $\beta_{\before{l}} = (\beta_{\before{l'}} - \setof{v}) \cup \variables{e}$. \\
Then $\beta_{\before{l}} \supseteq \beta_{\before{l'}} - \setof{v}$ and $\beta_{\before{l}} \cup \setof{v} \supseteq \beta_{\before{l'}}$.
}

\item By program execution rule for $l:v:=e \in P$, with $l' = \mathbf{next}(l)$, $\sigma' = \sigma[v \mapsto n]$,  \\
$\pi = \beta_{\before{l}}$,  and $\pi' = \beta_{\before_{l'}}$, 
$\tuple{l, \sigma, \beta_{\before{l}}} \Rightarrow \tuple{l', \sigma', \beta_{\before{l'}}}$.
\end{itemize}

\item $l:c = l:\mathbf{if}\; b \; \mathbf{then}\; g \in P$: 
\begin{itemize}
  \item Facts from dataflow equations: \\
    $\beta_{\before{l}} = \beta_{\after{l}} \cup \variables{b}$ (from transfer function $f$ for $l:\mathbf{if}\; b \; \mathbf{then}\; g \in P$) and \\
    $\beta_{\after{l}} \supseteq \beta_{\before{l'}}$ (because $l' \in \successors{l}$). 

  \item By Lemma~\ref{lem:lvbaee}, $\tuple{l, \sigma} \rightarrow \tuple{l', \sigma'}$ and $\variables{e} \subseteq \beta_{\before{l}}$ 
    imply $\tuple{b,\sigma,\beta_{\before{l}}} \Rightarrow t$, where $\tuple{b,\sigma} \rightarrow t$.

  \item Prove prophecy variable precondition $\forall v \in \variables{b} . v \in \beta_{\before{l}}$, i.e., prove $\variables{b} \subseteq \beta_{\before{l}}$: \\
    $\variables{b} \subseteq \beta_{\after{l}} \cup \variables{b} = \beta_{\before{l}}$

    \item Prove consistent with prophecy variable prediction $\beta_{\before{l'}} \subseteq \beta_{\before{l}}$: 
     Lemma~\ref{lem:pvconsistentsubseteq}.
\comment{
     Because $\beta_{\after{l}} \supseteq \beta_{\before{l'}}$, $\beta_{\before{l'}} \subseteq \beta_{\before{l}}$.
}

    \item By program execution rule for $l:\mathbf{if}\; b \; \mathbf{then}\; g \in P$ with $l'=g$ if $\sigma(b) = \textbf{true}$ or $l'=\textbf{next}(l)$ if $\sigma(b) = \textbf{false}$, $\sigma' = \sigma$, 
$\pi= \beta_{\before{l}} \supseteq \beta_{\before{l'}}$, and 
$\tuple{l, \sigma, \beta_{\before{l}}} \Rightarrow \tuple{l', \sigma', \beta_{\before{l'}}}$.
\end{itemize}

\item $l:c = l:\mathbf{goto}\; g \in P$, $l:c = l:\mathbf{skip} \in P$, or $l:c = l:\mathbf{halt} \in P$:
\begin{itemize}

	\item Facts from dataflow equations: \\
	$\beta_{\before{l}} = \beta_{\after{l}}$ (from transfer function $f$ for  
        $l:\mathbf{goto}\; g \in P$, $l:\mathbf{skip} \in P$, or $l:\mathbf{halt} \in P$) and \\
	$\beta_{\after{l}} = \beta_{\before{l'}}$ (because $\setof{l'} = \successors{l}$). 

	\item Prove prophecy variable precondition: \\
	There is no prophecy variable precondition for 
        $l:\mathbf{goto}\; g \in P$, $l:\mathbf{skip} \in P$, or $l:\mathbf{halt} \in P$. 

	\item Prove consistent with prophecy variable prediction $\beta_{\before{l'}} = \beta_{\before{l}}$: Lemma~\ref{lem:pvconsistenteq}.

\comment{
        $\beta_{\before{l}} = \beta_{\after{l}}$ and $\beta_{\after{l}} = \beta_{\before{l'}}$ imply $\beta_{\before{l}} = \beta_{\before{l'}}$.
}
	\item By the program execution rule for:
	\begin{itemize}
		\item $l:\mathbf{goto}\; g \in P$ with $l'=g$, $\sigma' =\sigma$, and 
                $\pi' = \beta_{\before{l'}} = \beta_{\before{l}} = \pi$, $\tuple{l, \sigma, \beta_{\before{l}}} \Rightarrow \tuple{l', \sigma', \beta_{\before{l'}}}$.
		\item $l:\mathbf{skip} \in P$ with $l'=\textbf{next}(l)$, $\sigma' =\sigma$, and 
                $\pi' = \beta_{\before{l'}} = \beta_{\before{l}} = \pi$, $\tuple{l, \sigma, \beta_{\before{l}}} \Rightarrow \tuple{l', \sigma', \beta_{\before{l'}}}$.
		\item $l:\mathbf{halt} \in P$ with $l'=\textbf{next}(l)$, $\sigma' =\sigma$, and 
                $\pi' = \beta_{\before{l'}} = \beta_{\before{l}} = \pi$, $\tuple{l, \sigma, \beta_{\before{l}}} \Rightarrow \tuple{l', \sigma', \beta_{\before{l'}}}$. Note that $\tuple{l,\sigma} \rightarrow \tuple{l',\sigma'}$ implies $\mathbf{next}(l) : \mathbf{done} \in P$. 
	\end{itemize}
\end{itemize}
\end{itemize}
\noindent$\blacksquare$

\subsubsection{Live Variables Correctness and Optimization Theorems} ~\label{sec:lvacot}

We next state and prove Theorem~\ref{thm:lvcharacterize}, which characterizes a relationship between the 
analysis and the program execution. Specifically, the theorem states that if a variable $v$ is not
live at some point in the execution and it is read in some future point in the execution, then
there is an intervening write to $v$ before it is read. 
Theorem~\ref{thm:lvcharacterize} ensures, for example, that if $v$ is not live 
immediately after an assigment (i.e., $l:v:=e \in P$ and $v \not\in \beta_{\after{l}}$,
it is possible to remove $l:v := e \in P$ without changing the result that the computation
produces. Note that Theorem~\ref{thm:lvcharacterize}
leverages the Progress theorem (Theorem~\ref{thm:lvprogress}) to use 
forward reasoning even though the key live variable properties deal with information about future program
executions and the analysis itself is a backward analysis. 

\begin{theorem}
\label{thm:lvcharacterize}
If $\tuple{l_i, \sigma_i} \rightarrow \cdots \rightarrow \tuple{l_j, \sigma_j}$, 
$v \not\in \beta_{\before{l_i}}$, $v \in \variables{c}$ where $l_j:c \in P$, 
then $\exists i \leq k < j . l_k:v:=e' \in P$:
\end{theorem}
\noindent{\bf Proof:} Find a $k$ that satisfies the theorem. \\
By Lemma~\ref{lem:lvvin}, $\variables{c} \subseteq \beta_{\before{l_j}}$. Then $v \in \variables{c}$ implies $v \in \beta_{\before{l_j}}$. \\
$v \not\in \beta_{\before{l_i}}$ and $v \in \beta_{\before{l_j}}$ imply $\exists i \leq k < j . v \not\in \beta_{l_k}$ 
and $v \in \beta_{\before{l_{k+1}}}$. \\
By Progress (Theorem~\ref{thm:vbeprogress}),
$\tuple{l_k, \sigma_k, \beta_{\before{l_k}}} \Rightarrow \tuple{l_{k+1}, \sigma_{k+1}, \beta_{\before{l_{k+1}}}}$. \\
By Lemma~\ref{lem:lvvup} $l_k : v := e' \in P$. 
\noindent$\blacksquare$

\subsubsection{Live Variables with Downward Closure Metarule}

The augmented operational semantics in Section~\ref{sec:lvaos} does not use the 
downward closure metarule (Definition~\ref{def:dcm}). It is possible to formulate
the analysis using this rule. Starting with the baseline augmented operational
semantics from Section~\ref{sec:baos}, update 
the rule for $l:v:=e \in P$ so that the prophecy variable $\pi$ predicts $v$ to be live 
after the assignment. The downward inference metarule then implements any
predictions that a variable $v$ becomes not live (specifically by removing
$v$ from $\pi$ after the execution of a command). 

\[
\infer{\tuple{l,\sigma, \pi}\Rightarrow \tuple{\mathbf{next}(l), \sigma[v\mapsto n], \pi \cup \setof{v}}}
      {l:v:=e\;\in\; \mathbf{P}\;\;\tuple{a,\sigma, \pi}\Rightarrow n}
\]

\noindent The only other update is to update the variable reference rule to include the prophecy variable precondition:

\[
\infer{\tuple{v,\sigma,\pi} \Rightarrow \sigma(v)}
      {v \in \dom{\sigma}\;\;\; v \in \pi}
\]

With this change the only difference between the augmented semantics with and without the
downward closure metarule is that, with the downward closure metarule, the prophecy
variable $\pi$ can predict that any command, 
including an $l:\mathbf{goto}\; g \in P$, $l:\mathbf{skip} \in P$ or $l:\mathbf{halt}\in P$ command, 
may transition a variable $v$ from live ($v \in \pi$) to not live ($v \not\in \pi'$). 
Without the downward closure metarule, only an $l:v:=e$ or $l:\mathbf{if}\; b \; \mathbf{then}\; g$ 
command can transition a variable $v$ from live ($v \in \pi$) to not live ($v \not\in \pi'$).
With the downward closure rule, the proofs in Sections~\ref{sec:lvaos} through \ref{sec:lvacot} go through
unchanged --- the only difference is that, in the Progress proof (Theorem~\ref{thm:lvprogress}), the dataflow facts for commands
$l:v:=e \in P$, $l:\mathbf{goto}\; g \in P$, $l:\mathbf{skip} \in P$ or $l:\mathbf{halt}\in P$
include $\beta_{\after{l}} \supseteq \beta_{\before{l'}}$ instead of $\beta_{\after{l}} = \beta_{\before{l'}}$,
so the proofs for these commands leverage a different case of Lemma~\ref{lem:pvconsistenteq}.

\subsection{Very Busy Expressions}~\label{sec:vbe}

The very busy expressions analysis (conservatively) determines, for each program point,
the expressions that are {\em very  busy} at that program point, i.e., expressions $e$ that,
in every terminating execution, must be evaluated before some $v \in \variables{e}$ is written. 
The analysis augments the standard operational semantics with a prophecy
variable $\pi \subseteq E$ that predicts which expressions are very busy. 
The prophecy variable $\pi$ is drawn from the 
very busy expressions program analysis lattice $\tuple{\Pi,\supseteq}$,
where $\Pi = \mathcal{P}(E)$, ordered
under reverse subset inclusion ($\supseteq$), with least upper bound $\cap$ and greatest lower bound
$\cup$. We use the notation $\subexpressions{e}$ is the set of subexpressions in $e$, 
$\subexpressions{b}$ is the set of subexpressions in $b$, and 
and $\subexpressions{e}$ is the set of subexpressions in $c$, 
all defined recursively over the structure of $e$, $b$, or $c$. 

\subsubsection{Augmented Operational Semantics} 
\label{sec:vbeos}

Starting with the baseline augmented operational semantics (Section~\ref{sec:baos}),
the analysis updates the program execution rule for $l:v:=e \in P$ to 
include a prophecy variable precondition to check for incorrect very busy
expression predictions. Specifically, the check 
requires that the predicted very busy expressions $\pi$ can contain an expression $e'$ only 
if 1) $e'$ is evaluated during the evaluation of $e$, i.e., $e' \in \subexpressions{e}$
or 2) $e'$ does not read $v$, i.e., $v \not\in \subexpressions{e'}$. The prophecy variable
then predicts that some new set $\pi'$ of expressions will be very busy after the execution
of $l:v:=e \in P$, with the constraint that $\pi'$ must include all previously predicted
very busy expressions not evaluated during the evaluation 
of $e$, i.e., $\pi' \supseteq \pi - \subexpressions{e}$. This last condition reflects
the fact that any expression that is very busy before $l:v:=e \in P$
and not evaluated by $e$ during the execution of $l:v:=e \in P$
must also be very busy after $l:v:=e \in P$. 

\[
\infer{\tuple{l,\sigma, \pi}\Rightarrow \tuple{\mathbf{next}(l), \sigma[v\mapsto n], \pi'}}
{l:v:=e\;\in\; \mathbf{P}\;\;\tuple{a,\sigma, \pi}\Rightarrow n\;\;\; 
e' \in \pi \;\mathbf{implies}\ e' \in \subexpressions{e}\; \mathbf{or}\; v \not\in \subexpressions{e'}\;\;\; \pi' \supseteq \pi - \subexpressions{e}
}
\]

The analysis similarly updates the rules for $l: \mathbf{if}\; b \; \mathbf{then} \; g \in P$ to predict 
new very busy expressions $\pi' \supseteq \pi - \subexpressions{b}$. Note that the predicted very busy expressions may increase
after the execution of $l: \mathbf{if}\; b \; \mathbf{then} \; g \in P$ because of the control flow
split --- there may be more very busy expressions after the split than before (because there
are fewer paths after the split than before) and different very busy expressions along the 
different control flow paths. 

\[
\infer{\tuple{l,\sigma, \pi}\Rightarrow \tuple{\mathbf{next}(l), \sigma, \pi'}}
{l\;:\;\mathbf{if}\;b\;\mathbf{then}\;g\;\in\; \mathbf{P}\;\;\; \tuple{b,\sigma, \pi} \Rightarrow \mathbf{false}\;\;\; \pi' \supseteq \pi - \subexpressions{b}}
\]

\[
\infer{\tuple{l,\sigma, \pi}\Rightarrow \tuple{g, \sigma, \pi'}}
{l:\mathbf{if}\;b\;\mathbf{then}\;g\;\in\; \mathbf{P}\;\;\; \tuple{b,\sigma, \pi} \Rightarrow \mathbf{true} \;\;\; \pi' \supseteq \pi - \subexpressions{b}}
\]

Finally the analysis also introduces a prophecy variable precondition for 
$l : \mathbf{halt}$ that requires that $\pi = \emptyset$, i.e., that
there are no predicted very busy expressions $\pi$ when the program halts:

\[
\infer{ 
\tuple{l, \sigma, \pi} \Rightarrow \tuple{\mathbf{next}(l), \sigma, \pi}
}{
l : \mathbf{halt} \in P \;\;\; \mathbf{next}(l):\mathbf{done} \in P \;\;\; \pi = \emptyset
}
\]

Unlike the live variables analysis, the very busy expressions analysis introduces no prophecy variable preconditions
into the expression evaluation rules, so expression evaluation is identical in the
standard and augmented semantics. 

\begin{lemma}
\label{lem:vbeee}
If $\tuple{e, \sigma, \pi} \Rightarrow n$, then $\tuple{e, \sigma} \rightarrow n$.
If $\tuple{b, \sigma, \pi} \Rightarrow t$, then $\tuple{b, \sigma} \rightarrow t$. \\
If $\tuple{e, \sigma} \rightarrow n$, then $\tuple{e, \sigma, \pi} \Rightarrow n$.
If $\tuple{b, \sigma} \rightarrow t$, then $\tuple{b, \sigma, \pi} \Rightarrow t$.
\end{lemma}
\noindent{\bf Proof:} The updated expression evaluation rules in the 
augmented operational semantics have the same preconditions 
and produce the same expression values as the corresponding rules from the standard operational semantics. 
\noindent$\blacksquare$

The following lemma characterizes the conditions under which an expression
$e'$ may leave the set of predicted very busy expressions $\pi$, specifically 
when the expression $e'$ is evaluated during the execution of a command
$l:c \in P$. The only commands $l:c \in P$ that remove expressions $e'$ from the
predicted very busy expressions $\pi$ (i.e., commands for which $\pi \subseteq \pi$)
are $l:v:=e \in P$ and $l: \mathbf{if}\; b \; \mathbf{then} \; g \in P$).

\begin{lemma} 
\label{lem:vbelem}
If $\tuple{l,\sigma,\pi} \Rightarrow \tuple{l', \sigma', \pi'}$, 
$e' \in \pi$, and $e' \not\in \pi'$, then $e' \in \subexpressions{c}$ where $l:c \in P$. 
\end{lemma}
\noindent{\bf Proof:} 
Case analysis on $l:c\in P$:
\begin{itemize}

\item $l:c = l:v:=e \in P$: 
In this case $\pi' \supseteq \pi - \subexpressions{e}$, so 
$e' \in \pi$ and $e' \not\in \pi$' implies $e' \in \subexpressions{e}$.

\item $l:c = l:\mathbf{if}\; b \; \mathbf{then}\; g \in P$: 
In this case $\pi' \supseteq \pi - \subexpressions{b}$, so 
$e' \in \pi$ and $e' \not\in \pi'$ implies $e' \in \subexpressions{b}$.

\item $l:c = l:\mathbf{skip} \in P$ or $l:c = l:\mathbf{goto}\; g \in P$: 
In this case $\pi = \pi'$, so there is no $e'$ such that $e' \in \pi$ and
$e' \not\in \pi'$.
\end{itemize}

\begin{theorem} 
	\label{thm:vbepreservation}
	(Preservation)
	If $\tuple{l, \sigma, \pi} \Rightarrow \tuple{l', \sigma', \pi'}$ then 
	$\tuple{l, \sigma} \rightarrow \tuple{l', \sigma'}$. 
\end{theorem}
\noindent{\bf Proof:} All rules that generate a transition
$\tuple{l, \sigma, \pi} \Rightarrow \tuple{l', \sigma', \pi'}$
in the augmented operational semantics
have the same preconditions over $l:c \in P$ and $\sigma$ and 
produce the same values for $l'$ and $\sigma'$ as the
corresponding rules from the standard operational semantics. 
\noindent$\blacksquare$

\subsubsection{Very Busy Expressions Analysis}

For each command $l:c \in P$, the analysis produces $\beta_{\before{l}} \subseteq E$ (the set of
very busy expressions at the program point before $l:c \in P$) and 
$\beta_{\after{l}} \subseteq E$ (the set of very busy expressions after $l:c \in P$). 
The analysis obtains the $\beta_{\before{l}}$ and $\beta_{\after{l}}$ 
by formulating and solving, using standard least fixed-point
techniques, the following set of backward dataflow equations:
\[
\begin{array}{rcl}
\beta_{\after{l}} & = & \emptyset \mbox{ if } l:\mathbf{done} \in P \\
\beta_{\after{l}} & = &  \cap \beta_{\before{g}}, \mbox{ where } l:c \in P \mbox{ and } g \in \successors{l} \\
\beta_{\before{l}} & = &  f(l, \beta_{\after{l}}) \\
\end{array}
\]
\noindent where $f$ is the transfer function for the analysis defined as follows:

\[
f(l, \beta) = 
\begin{cases}
(\beta - \setof{e' \in \beta . v \in \variables{e'}}) \cup \subexpressions{e} & \mbox{if }  l:v:=e \in P \\
\beta \cup \subexpressions{b} & \mbox{if }  l:\; \mbox{if}\; b \; \mbox{then}\; l' \\
\beta & \mbox{otherwise}
\end{cases}
\]

The next lemma states that if a variable $v$ is one of the variables referenced in an expression $e'$,
then the analysis determines that $e'$ is not very busy before an 
assignment $l:v:=e$ unless $e'$ is evaluated as part of the evaluation of $e$:
\begin{lemma}
\label{lem:vbeenotin}
If $l:v:= e \in P$, $v \in \variables{e'}$, and $e' \not\in \subexpressions{e}$,
then $e' \not\in \beta_{\before{l}}$. 
\end{lemma}
\noindent{\bf Proof:} 
By definition of transfer function $f$ for $l:v:= e \in P$, \\
$\beta_{\before{l}} = (\beta_{\after{l}} - \setof{e'\in\beta_{\before{l}} . v \in \variables{e'}}) \cup \subexpressions{e})$. 
Then if $v \in \variables{e'}$ and $e' \not\in \subexpressions{e}$, then $e' \not\in \beta_{\before{l}}$. 
\noindent$\blacksquare$

\subsubsection{Prophecy Variable Predictions and Dataflow Analyses with Sets of Elements and Reverse Subset Inclusion}

Very busy expressions is an instance of a more general class of backward dataflow analyses
in which the dataflow facts $\beta \in \Pi$ are sets of elements
with the dataflow lattice $\tuple{\Pi, \leq}$ ordered by reverse subset inclusion ($\supseteq$),
with least upper bound $\cap$, greatest lower bound $\cup$, and dataflow equations of the following form:
\[
\begin{array}{rcl}
\beta_{\after{l}} & = &  \cap \beta_{\before{g}}, \mbox{ where } l:c \in P \mbox{ and } g \in \successors{l} \\
\beta_{\before{l}} & = &  (\beta_{\after{l}} - D_l) \cup U_l
\end{array}
\]
\noindent Note that these equations ensure $\beta_{\before{l'}} \supseteq \beta_{\after{l}}$ for $l' \in \successors{l}$.
For very busy expressions $D_l = \setof{e' \in \beta . v \in \variables{e'}}$ and
$U_l = \subexpressions{e}$ when $l:v:=e \in P$; $D_l = \emptyset$ and $U_l = \subexpressions{b}$ when
$l:\; \mbox{if}\; b \; \mbox{then}\; l' \in P$. For $l:c = l:\mathbf{goto}\; g \in P$,
$l:c = l:\mathbf{skip} \in P$, or $l:c = l:\mathbf{halt} \in P$, $D_l = U_l = \emptyset$. 

We next show that prophecy variable predictions $\pi' \supseteq \pi - U_l$ are consistent with the
results that these analyses produce:
\begin{lemma}
\label{lem:pvconsistentsupseteq}
If $\beta_{\before{l}} = (\beta_{\after{l}} - D_l) \cup U_l$ and $\beta_{\before{l'}} \supseteq \beta_{\after{l}}$, 
then $\beta_{\before{l'}} \supseteq \beta_{\before{l}} - U_l$. 
\end{lemma}
\noindent{\bf Proof:} 
\begin{itemize}
\item  Known facts from dataflow analysis:
$\beta_{\before{l}} = (\beta_{\after{l}} - D_l) \cup U_l$ and $\beta_{\before{l'}} \supseteq \beta_{\after{l}}$.
\item Then $\beta_{\before{l}} \subseteq (\beta_{\before{l'}} - D_l) \cup U_l$, $\beta_{\before{l}} \subseteq \beta_{\before{l'}} \cup U_l$, 
$\beta_{\before{l}} - U_l \subseteq (\beta_{\before{l'}} \cup U_l) - U_l$, 
$\beta_{\before{l}} - U_l \subseteq \beta_{\before{l'}} - U_l$, and 
$\beta_{\before{l}} - U_l \subseteq \beta_{\before{l'}}$.
\end{itemize}
\noindent$\blacksquare$

If $D_l = U_l = \emptyset$ so that $f(l,\beta) = \beta$ and $\setof{l'} = \successors{l}$,
then $\beta_{\before{l}} = \beta_{\after{l}} = \beta_{\before{l'}}$. For live 
variables this is the case for $l:c = l:\mathbf{goto}\; g \in P$,
$l:c = l:\mathbf{skip} \in P$, and $l:c = l:\mathbf{halt} \in P$. 
In this case the analysis results are consistent with prophecy
variable predictions $\pi' = \pi$ and $\pi' \supseteq \pi$
(as in, for example, analyses that use the downward closure metarule):
\begin{lemma}
\label{lem:pvconsistenteqsupset}
If $\beta_{\before{l}} = \beta_{\after{l}}$ and $\beta_{\after{l}} = \beta_{\before{l'}}$, then
$\beta_{\before{l}} = \beta_{\before{l'}}$ and $\beta_{\before{l'}} \supseteq \beta_{\before{l}}$. 
\end{lemma}
\noindent$\blacksquare$

\subsubsection{Very Busy Expressions Progress Theorem}

We next state and prove the Progress theorem for the very busy expressions analysis. 
As in the Live Variables Progress theorem (Theorem~\ref{thm:lvprogress}), the prophecy variable
preconditions apply to $\beta_{\before{l}}$ and are immediately satisfied 
by the transfer function $f$ regardless of the values of $\beta_{\after{l}}$ and $\beta_{\before{l'}}$. 
As in the Live Variables Progress theorem (Theorem~\ref{thm:lvprogress}), 
the proof discharges the proof obligations required to show that the
analysis results $\beta_{\before{l}}$, $\beta_{\after{l}}$, and $\beta_{\before{l'}}$ are consistent with 
the prophecy variable predictions by pushing the analysis result
$\beta_{l'}$ through the transfer function for $l:c \in P$.

Because the analysis and prophecy variable predictions conform to the 
requirements of Lemmas~\ref{lem:pvconsistentsupseteq} and \ref{lem:pvconsistenteqsupset}, 
the prophecy variable prediction proof obligations for these commands are 
immediately discharged by applying these lemmas. 

\begin{theorem}
	\label{thm:vbeprogress}
	(Progress)
	If $\tuple{l, \sigma} \rightarrow \tuple{l', \sigma'}$ then 
	$\tuple{l, \sigma, \beta_{\before{l}}} \Rightarrow \tuple{l', \sigma', \beta_{\before{l'}}}$.
\end{theorem}

\noindent{\bf Proof:} 
If all of the prophecy variable preconditions are satisfied, 
the standard and augmented program execution rules for $l:c \in P$ define the
same values for $l'$, $\sigma'$, and all evaluated expressions $e$ or $b$. 
The following case analysis on $l:c \in P$ shows that the prophecy variable 
preconditions are satisfied and that 
$\beta_{\before{l}}$ and $\beta_{\before{l'}}$ satisfy 
$\tuple{l, \sigma, \beta_{\before{l}}} \Rightarrow \tuple{l', \sigma', \beta_{\before{l'}}}$:

\begin{itemize}

\item $l:c = l:v:=e \in P$: 
\begin{itemize}
\item Facts from dataflow equations: \\
$\beta_{\before{l}} = (\beta_{\after{l}} - \setof{e' \in \beta_{\after{l}} . v \in \variables{e'}}) \cup \subexpressions{e}$
(from transfer function $f$ for $l:v:=e \in P$) and 
$\beta_{\after{l}} = \beta_{\before{l'}}$ (because $\setof{l'} = \successors{l})$. 

\item By Lemma~\ref{lem:vbeee}, $\tuple{l, \sigma} \rightarrow \tuple{l', \sigma'}$ implies 
$\tuple{e,\sigma,\beta_{\before{l}}} \Rightarrow n$, where $\tuple{e,\sigma} \rightarrow n$.

\item Prove prophecy variable precondition 
$e' \in \beta_{\before{l}} \;\mathbf{implies}\ e' \in \subexpressions{e}\; \mathbf{or}\; v \not\in \subexpressions{e'}$: \\
Consider any $e' \in \beta_{\before{l}}$. 
Because $\beta_{\before{l}} = (\beta_{\after{l}} - \setof{e' \in \beta_{\after{l}} . v \in \variables{e'}}) \cup \subexpressions{e}$, 
either $e' \in \subexpressions{e}\; \mathbf{or}\; v \not\in \subexpressions{e'}$.

\item Prove consistent with prophecy variable prediction 
$\beta_{\before{l'}} \supseteq \beta_{\before{l}} - \subexpressions{e}$: Lemma~\ref{lem:pvconsistentsupseteq}.
\comment{
Because 
$\beta_{\before{l}} = (\beta_{\after{l}} - \setof{e'\in\beta_{\before{l}} . v \in \variables{e'}}) \cup \subexpressions{e})$
and $\beta_{\after{l}} = \beta_{\before{l'}}$, \\
$\beta_{\before{l}} = (\beta_{\before{l'}} - \setof{e'\in\beta_{\before{l}} . v \in \variables{e'}}) \cup \subexpressions{e})$. \\
Then $\beta_{\before{l}} \supseteq \beta_{\before{l'}} \cup \subexpressions{e})$ and $\beta_{\before{l}} - \subexpressions{e} \supseteq \beta_{\before{l'}}$.
}

\item By program execution rule for $l:v:=e \in P$, with $l' = \mathbf{next}(l)$, $\sigma' = \sigma[v \mapsto n]$,  \\
$\pi = \beta_{\before{l}}$,  and $\pi' = \beta_{\before_{l'}}$, 
$\tuple{l, \sigma, \beta_{\before{l}}} \Rightarrow \tuple{l', \sigma', \beta_{\before{l'}}}$.
\end{itemize}

\item $l:c = l:\mathbf{if}\; b \; \mathbf{then}\; g \in P$: 
\begin{itemize}
  \item Facts from dataflow equations: \\
    $\beta_{\before{l}} = \beta_{\after{l}} \cup \subexpressions{b}$
    (from transfer function $f$ for $l:\mathbf{if}\; b \; \mathbf{then}\; g \in P$) and \\
    $\beta_{\after{l}} \subseteq \beta_{\before{l'}}$ (because $l' \in \successors{l})$. 

  \item By Lemma~\ref{lem:vbeee}, $\tuple{l, \sigma} \rightarrow \tuple{l', \sigma'}$ implies 
    $\tuple{b,\sigma,\beta_{\before{l}}} \Rightarrow t$, where $\tuple{e,\sigma} \rightarrow t$.

   \item Prove prophecy variable precondition: \\
          There is no prophecy variable precondition for $l:\mathbf{if}\; b \; \mathbf{then}\; g \in P$.

   \item Prove consistent with prophecy variable prediction 
     $\beta_{\before{l'}} \supseteq \beta_{\before{l}} - \subexpressions{b}$: Lemma~\ref{lem:pvconsistentsupseteq}.
\comment{
     Because $\beta_{\before{l}} = \beta_{\after{l}} \cup \subexpressions{b}$ and $\beta_{\after{l}} \subseteq \beta_{\before{l'}}$, 
     $\beta_{\before{l}} \subseteq \beta_{\before{l'}} \cup \subexpressions{b}$. Then 
     $\beta_{\before{l}} - \subexpressions{b} \subseteq \beta{\before{l'}}$. 
}

    \item By program execution rule for $l:\mathbf{if}\; b \; \mathbf{then}\; g \in P$ with $l'=g$ if $\sigma(b) = \textbf{true}$ or $l'=\textbf{next}(l)$ if $\sigma(b) = \textbf{false}$, $\sigma' = \sigma$, 
$\pi= \beta_{\before{l}} \supseteq \beta_{\before{l'}} \cup \subexpressions{b}$, and 
$\pi' = \beta_{\before{l'}}$,
$\tuple{l, \sigma, \beta_{\before{l}}} \Rightarrow \tuple{l', \sigma', \beta_{\before{l'}}}$.
\end{itemize}

\item $l:c = l:\mathbf{goto}\; g \in P$ or $l:c = l:\mathbf{skip} \in P$:
\begin{itemize}
	\item Facts from dataflow equations: \\
	$\beta_{\before{l}} = \beta_{\after{l}}$ (from transfer function $f$ for  
        $l:\mathbf{goto}\; g \in P$, $l:\mathbf{skip} \in P$, or $l:\mathbf{halt} \in P$) and \\
	$\beta_{\after{l}} = \beta_{\before{l'}}$ (because $\setof{l'} = \successors{l}$). 

	\item Prove prophecy variable precondition: \\
	There is no prophecy variable precondition for 
        $l:\mathbf{goto}\; g \in P$, $l:\mathbf{skip} \in P$, or $l:\mathbf{halt} \in P$. 

	\item Prove consistent with prophecy variable prediction $\beta_{\before{l'}} = \beta_{\before{l}}$: Lemma~\ref{lem:pvconsistenteqsupset}.

\comment{
        $\beta_{\before{l}} = \beta_{\after{l}}$ and $\beta_{\after{l}} = \beta_{\before{l'}}$ imply $\beta_{\before{l}} = \beta_{\before{l'}}$.
}

	\item By the program execution rule for:
	\begin{itemize}
		\item $l:\mathbf{goto}\; g \in P$ with $l'=g$, $\sigma' =\sigma$, and 
                $\pi' = \beta_{\before{l'}} = \beta_{\before{l}} = \pi$, $\tuple{l, \sigma, \beta_{\before{l}}} \Rightarrow \tuple{l', \sigma', \beta_{\before{l'}}}$.
		\item $l:\mathbf{skip} \in P$ with $l'=\textbf{next}(l)$, $\sigma' =\sigma$, and 
                $\pi' = \beta_{\before{l'}} = \beta_{\before{l}} = \pi$, $\tuple{l, \sigma, \beta_{\before{l}}} \Rightarrow \tuple{l', \sigma', \beta_{\before{l'}}}$.
		\item $l:\mathbf{halt} \in P$ with $l'=\textbf{next}(l)$, $\sigma' =\sigma$, and 
                $\pi' = \beta_{\before{l'}} = \beta_{\before{l}} = \pi$, $\tuple{l, \sigma, \beta_{\before{l}}} \Rightarrow \tuple{l', \sigma', \beta_{\before{l'}}}$. Note that $\tuple{l,\sigma} \rightarrow \tuple{l',\sigma'}$ implies $\mathbf{next}(l) : \mathbf{done} \in P$. 
	\end{itemize}
\end{itemize}

\item $l:c = l:\mathbf{halt} \in P$:
\begin{itemize}
	\item Facts from dataflow equations: \\
	$\beta_{\before{l}} = \beta_{\after{l}} = \emptyset$ (from transfer function $f$ for $:\mathbf{halt} \in P$) and \\
	$\beta_{\after{l}} = \beta_{\before{l'}}$ (because $\setof{l'} = \successors{l}$). 
	\item Prove prophecy variable precondition $\beta_{\before{l}} = \emptyset$: \\
         Because $\beta_{\before{l}} = \beta_{\after{l}} = \emptyset$, $\beta_{\before{l}} = \emptyset$.
	\item Prove consistent with prophecy variable prediction $\beta_{\before{l'}} = \beta_{\before{l}}$: \\
        $\beta_{\before{l}} = \beta_{\after{l}}$ and $\beta_{\after{l}} = \beta_{\before{l'}}$ imply $\beta_{\before{l}} = \beta_{\before{l'}}$.
	\item By the program execution rule for $l:\mathbf{halt} \in P$ with $l'=\textbf{next}(l)$, $\sigma' =\sigma$, 
                $\pi' = \beta_{\before{l'}} = \beta_{\before{l}} = \pi$, $\tuple{l, \sigma, \beta_{\before{l}}} \Rightarrow \tuple{l', \sigma', \beta_{\before{l'}}}$. Note that $\tuple{l,\sigma} \rightarrow \tuple{l',\sigma'}$ implies $\mathbf{next}(l) : \mathbf{done} \in P$. 
\end{itemize}

\end{itemize}
\noindent$\blacksquare$

\subsubsection{Very Busy Expressions Correctness Theorems}

The following two theorems establish correctness properties of 
the very busy expressions analysis. The first states that,
in all executions, all very busy expressions $e'$ are evaluated before 
any of the variables $v \in \variables{e'}$ is reassigned. 
The second states that no execution halts before evaluating
all very busy expressions $e'$. These are the correctness properties required
to establish the soundness of, for example, standard code hoisting optimizations
that use very busy expressions~\cite{cooper2011engineering}.

\begin{theorem}
\label{thm:vbeca}
If $\tuple{l_i, \sigma_i} \rightarrow \cdots \rightarrow \tuple{l_j, \sigma_j}$, 
$l_j : v := e \in P$, $e' \in \beta_{\before{l_i}}$, $v \in \variables{e'}$, 
then $\exists i \leq k \leq j . e' \in \subexpressions{c}$ where $l_k:c \in P$. 
\end{theorem}
\noindent{\bf Proof:} Find a $k$ that satisfies the theorem. \\
If $e' \in \subexpressions{e}$, then $k = j$.  \\
If $e' \not\in \subexpressions{e}$, then by Lemma~\ref{lem:vbeenotin}, $e' \not\in \beta_{\before{l_j}}$. \\
$e' \in \beta_{\before{l_i}}$ and $e' \not\in \beta_{\before{l_j}}$ imply $\exists i \leq k < j . e' \in \beta_{l_k}$ 
and $e' \not\in \beta_{\before{l_{k+1}}}$. \\
By Progress (Theorem~\ref{thm:vbeprogress}),
$\tuple{l_k, \sigma_k, \beta_{\before{l_k}}} \Rightarrow \tuple{l_{k+1}, \sigma_{k+1}, \beta_{\before{l_{k+1}}}}$. \\
By Lemma~\ref{lem:vbelem} $e' \in \subexpressions{c}$ where $l_k:c \in P$. 
\noindent$\blacksquare$

\begin{theorem}~\label{thm:vbech}
If $\tuple{l_i, \sigma_i} \rightarrow \cdots \rightarrow \tuple{l_j,\sigma_j}$, $e' \in \beta_{l_i}$, and 
$l_j : \mathbf{done} \in P$, then $\exists i \leq k < j . e' \in \subexpressions{c}$ where $l:c \in P$. 
\end{theorem}
\noindent{\bf Proof:} Find a $k$ that satisfies the theorem. \\
$l_j : \mathbf{done} \in P$ implies $\beta_{\before{l_j}} = \beta_{\after{l_j}} = \emptyset$. Then  $e' \not\in \beta_{\before{l_j}}$. \\
$e' \in \beta_{\before{l_i}}$ and $e' \not\in \beta_{\before{l_j}}$ imply $\exists i \leq k < j . e' \in \beta_{l_k}$ 
and $e' \not\in \beta_{\before{l_{k+1}}}$. \\
By Progress (Theorem~\ref{thm:vbeprogress}),
$\tuple{l_k, \sigma_k, \beta_{\before{l_k}}} \Rightarrow \tuple{l_{k+1}, \sigma_{k+1}, \beta_{\before{l_{k+1}}}}$. \\
By Lemma~\ref{lem:vbelem} $e' \in \subexpressions{c}$ where $l_k:c \in P$. 
\noindent$\blacksquare$

\subsection{Prophecy Variables, Dataflow Analyses, and Complemented, Distributive Lattices}

We next generalize Lemmas~\ref{lem:pvconsistentsubseteq} and \ref{lem:pvconsistentsupseteq} to arbitrary 
complemented, distributive lattices $\tuple{\Pi, \leq}$ ordered by $\leq$, with least upper bound $\lor$, 
greatest lower bound $\land$, greatest element $\top$, and least element $\bot$. For each $\beta \in \Pi$,
$\overline{\beta}$ is the {\em complement} of $\beta$, i.e., the unique lattice element $\overline{\beta} \in \Pi$
such that $\beta \lor \overline{\beta} = \top$ and $\beta \land \overline{\beta} = \bot$. 
Note that if the $\beta \in \Pi$ are sets of elements from some underlying set $S$
with $\Pi = \mathcal{P}(S)$, then the lattice $\tuple{\Pi,\subseteq}$
and the lattice $\tuple{\Pi, \supseteq}$ are both complemented, distributive lattices 
where $\overline{\beta}$ is set complement; i.e. $\overline{\beta} = S - \beta$. For
$\tuple{\Pi, \subseteq}$, $\beta - D = \beta \land \overline{D}$. For $\tuple{\Pi, \supseteq}$, $\beta - D = \beta \lor \overline{D}$.

We consider two cases for arbitrary complemented, distributive lattices:
\[
\begin{array}{rcl}
\beta_{\after{l}} & = &  \lor \beta_{\before{g}}, \mbox{ where } l:c \in P \mbox{ and } g \in \successors{l} \\
\beta_{\before{l}} & = &  (\beta_{\after{l}} \land \overline{D_l}) \lor U_l
\end{array}
\]
\noindent where the analysis results are consistent with the prophecy variable prediction $\pi' \leq \pi \lor D_l$
and
\[
\begin{array}{rcl}
\beta_{\after{l}} & = &  \lor \beta_{\before{g}}, \mbox{ where } l:c \in P \mbox{ and } g \in \successors{l} \\
\beta_{\before{l}} & = &  (\beta_{\after{l}} \lor \overline{D_l}) \land U_l
\end{array}
\]
\noindent where the analysis results are consistent with the prophecy variable prediction $\pi' \leq \pi \lor \overline{U_l}$.

\begin{lemma}
\label{lem:pvconsistentsubseteqcd}
If $\beta_{\before{l}} = (\beta_{\after{l}} \land \overline{D_l}) \lor U_l$ and $\beta_{\before{l'}} \leq \beta_{\after{l}}$, 
then $\beta_{\before{l'}} \leq \beta_{\before{l}} \lor D_l$. 
\end{lemma}
\noindent{\bf Proof:} 
\begin{itemize}
\item  Known facts from dataflow analysis:
$\beta_{\before{l}} = (\beta_{\after{l}} \land \overline{D_l}) \lor U_l$ and $\beta_{\before{l'}} \leq \beta_{\after{l}}$.
\item Then 
$\beta_{\before{l}} \geq (\beta_{\before{l'}} \land \overline{D_l}) \lor U_l$,
$\beta_{\before{l}} \geq (\beta_{\before{l'}} \land \overline{D_l})$, 
$\beta_{\before{l}} \lor D_l \geq (\beta_{\before{l'}} \land \overline{D_l}) \lor D_l$, 
$\beta_{\before{l}} \lor D_l \geq (\beta_{\before{l'}} \lor D_l) \land (\overline{D_l} \lor D_l)$, 
$\beta_{\before{l}} \lor D_l \geq (\beta_{\before{l'}} \lor D_l) \land \top$, 
$\beta_{\before{l}} \lor D_l \geq (\beta_{\before{l'}} \lor D_l)$, and
$\beta_{\before{l}} \lor D_l \geq \beta_{\before{l'}})$.
\end{itemize}
\noindent$\blacksquare$

\begin{lemma}
\label{lem:pvconsistentsupseteqcd}
If $\beta_{\before{l}} = (\beta_{\after{l}} \lor \overline{D_l}) \land U_l$ and $\beta_{\before{l'}} \leq \beta_{\after{l}}$, 
then $\beta_{\before{l'}} \leq \beta_{\before{l}} \lor \overline{U_l}$. 
\end{lemma}
\noindent{\bf Proof:} 
\begin{itemize}
\item  Known facts from dataflow analysis:
$\beta_{\before{l}} = (\beta_{\after{l}} \lor \overline{D_l}) \land U_l$ and $\beta_{\before{l'}} \leq \beta_{\after{l}}$.
\item Then 
$\beta_{\before{l}} \geq (\beta_{\after{l}} \lor \overline{D_l}) \land U_l$,
$\beta_{\before{l}} \geq \beta_{\after{l}} \land U_l$,
$\beta_{\before{l}} \lor \overline{U_1} \geq (\beta_{\after{l}} \land U_l) \lor \overline{U_1}$,
$\beta_{\before{l}} \lor \overline{U_1} \geq (\beta_{\after{l}} \lor \overline{U_1}) \land (\overline{U_1} \lor U_l)$, 
$\beta_{\before{l}} \lor \overline{U_1} \geq (\beta_{\after{l}} \lor \overline{U_1}) \land \top$, 
$\beta_{\before{l}} \lor \overline{U_1} \geq (\beta_{\after{l}} \lor \overline{U_1})$, and 
$\beta_{\before{l}} \lor \overline{U_1} \geq \beta_{\after{l}}$.
\end{itemize}
\noindent$\blacksquare$

	\section{Forward Analyses and History Variables}
\label{sec:forward}

We next present several forward analyses that use history variables
to record information about the past execution of the program.

\subsection{Defined Variables Analysis}~\label{sec:dv}

The standard operational semantics in Figures~\ref{fig:eeer} -- \ref{fig:per} will become stuck if
an expression reads the value of an undefined variable. We next present an analysis 
that computes, for each program point, the variables that are 
defined on all program execution paths to that program point. This analysis can
be used to (conservatively) check if any program execution can become stuck
because it attempts to access an undefined variable. 

\subsubsection{Defined Variables Augmented Operational Semantics}

Starting with the baseline augmented operational semantics (Section~\ref{sec:baos}), 
the analysis augments the baseline semantics with a 
history variable $\pi \subseteq V$. $\pi$ records (a subset of) the variables $v \in V$ that are defined in each 
configuration $\tuple{l,\sigma,\pi}$. The program analysis lattice $\tuple{\Pi,\supseteq}$ is ordered
under reverse subset inclusion ($\supseteq$) with least upper bound $\cap$ and greatest lower bound $\cup$. 
The augmented operational semantics
for this analysis updates the program execution rule for commands $l : v := e \in P$ to update the 
history variable $\pi$ to record $v$ as one of the defined variables. 
All other rules remain unchanged and the analysis
applies the upward closure metarule (Definition~\ref{def:ucm}).
$\pi_0 = \emptyset$ is the initial value of the history variable $\pi$. 

\[
\infer{\tuple{l,\sigma,\pi}\Rightarrow \tuple{\mathbf{next}(l) ,\sigma[v\mapsto n], \pi \cup \setof{v}}}
		{l:v:=e\;\in\; \mathbf{P}\;\;\tuple{e,\sigma,\pi}\Rightarrow n}
\]

We next state and prove a lemma that the history variable $\pi$ records a subset of the
defined variables at every point in the execution. Without the upward closure metarule
$\pi = \dom{\sigma}$. The upward closure metarule enables the augmented operational
semantics to drop defined variables $v$ from $\pi$ so that $\pi \subseteq \dom{\sigma}$. 

\begin{lemma}
\label{lem:dv}
$\tuple{l_0, \sigma_0, \pi_0} \Rightarrow \cdots \rightarrow \tuple{l_i, \sigma_i, \pi_i}$
implies $\pi_i \subseteq \dom{\sigma_i}$.
\end{lemma}
\noindent{\bf Proof} (induction on $i$):\\
Base Case: ($i=0$): $\pi_0 = \emptyset \subseteq \dom{\sigma_0}$.  \\
Induction Step: (assume for $i$, prove for $i+1$): \\
$\tuple{l_i,\sigma_i, \pi_{i}}\Rightarrow\tuple{l_{i+1},\sigma_{i+1},\pi_{i+1}}$ \\
Case analysis on $l_i:c \in P$: 
\begin{itemize}
  \item $l_i:c = l_i:v:=e \in P$:
    \begin{itemize}
      \item Known facts:
      \begin{itemize}
        \item $\sigma_{i+1} = \sigma_i[v \mapsto n]$ where $\tuple{e,\sigma_i, \pi_i} \Rightarrow n$.
        \item $\pi_{i+1} \subseteq \pi_i \cup \setof{v}$. 
        \item $\pi_i \subseteq \dom{\sigma_i}$. 
      \end{itemize}
      \item Then $\dom{\sigma_{i+1}} = \dom{\sigma_i} \cup \setof{v}$ and $\pi_i \cup \setof{v} \subseteq \dom{\sigma_i} \cup \setof{v}$.
           So $\pi_i \cup \setof{v} \subseteq \dom{\sigma_{i+1}}$ and $\pi_{i+1} \subseteq \dom{\sigma_{i+1}}$. 
    \end{itemize}
   
\item $l_i:c = l_i:\mathbf{if}\; b \; \mathbf{then}\; g$, $l_i:c = l_i:\mathbf{goto}\; g$, 
  $l_i:c = l_i:\mathbf{skip}$, or $l_i:c = l_i:\mathbf{halt}$: \\
  Then $\pi_{i+1} = \pi_i$ and $\sigma_{i+1} = \sigma_i$.
  By induction hypothesis $\pi_i \subseteq \dom{\sigma_i}$, so $\pi_{i+1} \subseteq \dom{\sigma_{i+1}}$
\end{itemize}
\noindent$\blacksquare$

The Preservation theorem is straightforward as the defined variables analysis introduces
no rule preconditions or changes to $l$ or $\sigma$. 
\begin{theorem}
\label{thm:dvpreservation}
(Preservation): If $\tuple{l,\sigma,\pi}\Rightarrow\tuple{l',\sigma',\pi'}$, then
$\tuple{l,\sigma}\Rightarrow\tuple{l',\sigma'}$.
\end{theorem}
\noindent{\bf Proof:} All rules that generate a transition
$\tuple{l, \sigma, \pi} \Rightarrow \tuple{l', \sigma', \pi'}$
in the augmented operational semantics
have the same preconditions over $l:c \in P$ and $\sigma$ and 
produce the same values for $l'$ and $\sigma'$ as the
corresponding rules from the standard operational semantics. 
\noindent$\blacksquare$

\subsubsection{Defined Variables Analysis}
The defined variables analysis obtains the analysis
results $\beta_{\before{l}}$ and $\beta_{\after{l}}$ 
by formulating and solving, using standard least fixed-point
techniques, the following set of forward dataflow equations:
\[
\begin{array}{rcl}
\beta_{\before{\first{P}}} & = & \emptyset \\
\beta_{\before{l}} & = &  \cap \beta_{\after{g}}, \mbox{ where } l:c \in P \mbox{ and } g \in \pred{l} \\
\beta_{\after{l}} & = &  f(l, \beta_{\before{l}})
\end{array}
\]
\noindent where $f$ is the transfer function for the analysis defined as follows:

\[
f(l, \beta) = 
\begin{cases}
\beta \cup \setof{v} & \mbox{if }  l:v:=e \in P \\
\beta & \mbox{otherwise}
\end{cases}
\]

\subsubsection{Defined Variables Progress Theorem} We next state and prove the Progress theorem for the defined variables analysis. 

\begin{theorem}
\label{thm:dvprogress}
(Progress): If $\tuple{l,\sigma}\rightarrow\tuple{l',\sigma'}$, then
$\tuple{l,\sigma, \beta_{\before{l}}}\Rightarrow\tuple{l',\sigma',\beta_{\before{l'}}}$.
\end{theorem}

\noindent{\bf Proof:} 
The following case analysis on $l:c \in P$ shows that the analysis
results are consistent with the history variable updates. 
so that $\beta_{\before{l}}$ and $\beta_{\before{l'}}$ satisfy 
$\tuple{l, \sigma, \beta_{\before{l}}} \Rightarrow \tuple{l', \sigma', \beta_{\before{l'}}}$:

\begin{itemize}
	\item $l:c = l:v:=e \in P$: 
	\begin{itemize}
		\item Facts from dataflow equations: \\
		$\beta_{\before{l}} \cup \setof{v} = \beta_{\after{l}}$ (from transfer function $f$ for $l:v:=e \in P$) \\
		$\beta_{\after{l}} \supseteq \beta_{\before{l'}}$ (because $l \in \pred{l'}$)  
		\item Prove history variable consistency (prove $\beta_{\before{l}} \cup\{v\} \supseteq \beta_{\before{l'}}$): \\
		 $\beta_{\before{l}} \cup \setof{v} = \beta_{\after{l}}$ and $\beta_{\after{l}} \supseteq \beta_{\before{l'}}$ imply $\beta_{\before{l}} \cup\{v\} \supseteq \beta_{\before{l'}}$
		 \item By the program execution rule for $l:v:=e \in P$, with $l' = \mathbf{next}(l)$, $\sigma' = \sigma[v \mapsto n]$, $\pi = \beta_{\before{l}}$, and $\pi' = \beta_{\before{l'}}, \tuple{l, \sigma, \beta_{\before{l}}} \Rightarrow \tuple{l', \sigma', \beta_{\before{l'}}}$.
	\end{itemize}
	\item $l:c = l:\mathbf{if}\; b \; \mathbf{then}\; g \in P ,  l:\mathbf{skip} \in P, l:\mathbf{goto}\; g\in P$: 
	\begin{itemize}
		\item Facts from dataflow equations: \\
		$\beta_{\before{l}} = \beta_{\after{l}}$ (from transfer function $f$ for $l:c \in P$) \\
		$\beta_{\after{l}} \supseteq \beta_{\before{l'}}$ (because $l \in \pred{l'}$)  
		\item Prove history variable consistency (prove $\beta_{\before{l}} \supseteq \beta_{\before{l'}}$): \\
		$\beta_{\before{l}} = \beta_{\after{l}}$ and $\beta_{\after{l}} \supseteq \beta_{\before{l'}}$ imply $\beta_{\before{l}} \supseteq \beta_{\before{l'}}$
		\item By program execution rule for $l:\mathbf{if}\; b \; \mathbf{then}\; g \in P$:
	\begin{itemize}
		 \item if $\tuple{l,\sigma,b} \Rightarrow\textbf{true}, l' = g$, $\sigma' = \sigma$, $\pi = \beta_{\before{l}}$ and $\pi' = \beta_{\before{l'}}, \tuple{l, \sigma, \beta_{\before{l}}} \Rightarrow \tuple{l', \sigma', \beta_{\before{l'}}}$.
		 \item if $\tuple{l,\sigma,b} \Rightarrow\textbf{false}, l' = \mathbf{next}(l)$, $\sigma' = \sigma$, $\pi = \beta_{\before{l}}$ and $\pi' = \beta_{\before{l'}}, \tuple{l, \sigma, \beta_{\before{l}}} \Rightarrow \tuple{l', \sigma', \beta_{\before{l'}}}$.
	 \end{itemize}
	 	\item By program execution rule for $l:\mathbf{skip} \in P$, with  $l' = \mathbf{next}(l)$, $\sigma' = \sigma$, $\pi = \beta_{\before{l}}$, and $\pi' = \beta_{\before{l'}}$:
	 	 $\tuple{l, \sigma, \beta_{\before{l}}} \Rightarrow \tuple{l', \sigma', \beta_{\before{l'}}}$
	 	 	 	\item By program execution rule for $l:\textbf{goto}\;g \in P$, with  $l' = g$, $\sigma' = \sigma$, $\pi = \beta_{\before{l}}$, and $\pi' = \beta_{\before{l'}}, \tuple{l, \sigma, \beta_{\before{l}}} \Rightarrow \tuple{l', \sigma', \beta_{\before{l'}}}$
	\end{itemize}
\end{itemize}
\noindent$\blacksquare$

\subsubsection{Defined Variables Correctness Theorem} We now state and prove a correctness theorem for the defined variable analysis.
At a high level, this theorem states that the analysis (conservatively) computes 
an under approximation of the variables that are defined at any point in the execution
of the program $P$ --- if the analysis says that a variable is defined, then it is
defined in all executions. 

\begin{theorem}
\label{thm:dv}
$\tuple{l_0, \sigma_0} \rightarrow \cdots \rightarrow \tuple{l_i, \sigma_i}$
implies $\beta_{\before{l_i}} \subseteq \dom{\sigma_i}$.
\end{theorem}
\noindent{\bf Proof}: By Progress (Theorem~\ref{thm:dvprogress}), 
$\tuple{l_0, \sigma_0, \beta_{\before{l_0}}} \Rightarrow \cdots \Rightarrow \tuple{l_i, \sigma_i, \beta_{\before{l_i}}}$, 
where $\beta_{\before_{l_0}} = \pi_0 = \emptyset$. 
By Lemma~\ref{lem:dv}, $\beta_{\before_{l_i}} \subseteq \dom{\sigma_i}$. 
\noindent$\blacksquare$

\comment{
\begin{definition}
$v$ is defined at $l:c \in P$ if $v \in \beta_{\before{l}}$. 
$e$ is defined at $l:c \in P$ if $\variables{e} \subseteq \beta_{\before{l}}$. 
$b$ is defined at $l:c \in P$ if $\variables{b} \subseteq \beta_{\before{l}}$. 
\end{definition}
}

In the standard program execution semantics (Figures~\ref{fig:eeer}-\ref{fig:per}), a program
execution becomes stuck at $l$ if the evaluation of an expression $e$ in $l:v:=e \in P$ 
or $b$ in $l:\mathbf{if}\; b\; \mathbf{goto}\; g \in P$ attempts to 
read an undefined variable $v$ (i.e., a variable $v \not\in \dom{\sigma}$).
But if the analysis determines that all variables in $e$ or $b$ are defined,
then the execution will not become stuck at the evaluation of $e$ or $b$ because
of an attempt to read an undefined variable.  We formalize this reasoning as follows:
\begin{theorem}
\label{thm:dvenotstuck}
If $\tuple{l_0, \sigma_0} \rightarrow \cdots \rightarrow \tuple{l_i, \sigma_i}$ and 
$\variables{e} \subseteq \beta_{\before{l_i}}$, then $\tuple{e,\sigma_i} \rightarrow n$. 
\end{theorem}
\begin{theorem}
\label{thm:dvbnotstuck}
If $\tuple{l_0, \sigma_0} \rightarrow \cdots \rightarrow \tuple{l_i, \sigma_i}$ and 
$\variables{b} \subseteq \beta_{\before{l_i}}$, then $\tuple{b,\sigma_i} \rightarrow t$. 
\end{theorem}
\noindent{\bf Proof}: By Theorem~\ref{thm:dv}, $\beta_{\before{l_i}} \subseteq \dom{\sigma_i}$, so 
$\variables{e} \subseteq \beta_{\before{l_i}} \subseteq \dom{\sigma_i}$, which ensures that
all variables are defined during the evaluation of $e$. 
Similarly $\variables{b} \subseteq \beta_{\before{l_i}} \subseteq \dom{\sigma_i}$ ensures that
all variables are defined during the evaluation of $b$.
\noindent$\blacksquare$

\subsection{Reaching Definitions}~\label{sec:rd}

Reaching definitions is a classic program analysis used, for example, in constant propagation
and other compiler optimizations~\cite{cooper2011engineering}. The analysis augments the standard 
operational semantics with a history variable $\pi: V \rightarrow  \mathcal{P}(L)$. $\pi\in \Pi$ records the most recent definition of a given variable $v\in V$ by recording, for each variable $v$, the label $l$ 
of the most recent assignment to $v$. The program analysis lattice $\tuple{\Pi, \leq}$ is ordered under element-wise subset inclusion 
(i.e., $\pi_1 \leq \pi_2$ if $\forall v \in V . \pi_1(v) \subseteq \pi_2(v)$) with 
least upper bound $\lor$ (i.e., $\pi_1 \lor \pi_2 = \lambda v \in V. \pi_1(v) \cup \pi_2(v)$)
and greatest lower bound $\land$ (i.e., $\pi_1 \land \pi_2 = \lambda v \in V. \pi_1(v) \cap \pi_2(v)$).
The augmented operational semantics updates the program execution rule for commands $l:v := e \in P$ to record 
the fact that $l$ is the current definition of $v$ (i.e., $\pi(v) = \setof{l}$). 
All other rules remain unchanged and we apply the upward closure metarule. 
$\pi_0 =\lambda v . \emptyset$ is the initial value for $\pi$.

\[
\infer{\tuple{l,\sigma,\pi}\Rightarrow \tuple{\mathbf{next}(l) ,\sigma[v\mapsto n], \pi[v \mapsto \setof{l}]}}
{l:v:=e\;\in\; \mathbf{P}\;\;\tuple{e,\sigma,\pi}\Rightarrow n}
\]

\begin{theorem}
	\label{thm:rdpreservation}
	(Preservation): If $\tuple{l,\sigma,\pi}\Rightarrow\tuple{l',\sigma',\pi'}$, then
	$\tuple{l,\sigma}\Rightarrow\tuple{l',\sigma'}$.
\end{theorem}
\noindent{\bf Proof:} All rules that generate a transition
$\tuple{l, \sigma, \pi} \Rightarrow \tuple{l', \sigma', \pi'}$
in the augmented operational semantics
have the same preconditions over $l:c \in P$ and $\sigma$ and 
produce the same values for $l'$ and $\sigma'$ as the
corresponding rules from the standard operational semantics. 
\noindent$\blacksquare$

We next prove a lemma related to the relationship between reaching definitions and
the values recorded in the program execution states $\sigma$, specifically that if a
state $\sigma$ records $n$ as the value of $v$, then one of the labels recorded in the
corresponding history variable $\pi(v)$ is the label of an executed assignment statement that
assigned the value $n$ to $v$. Note that (in the absence of any defined variable information
as could be computed, for example, by the defined variables analysis from Section~\ref{sec:dv})
there is no guarantee that any executed assignment command $l:v:=e \in P$ assigned a value to $v$
and no guarantee that $v$ is defined.

\begin{lemma}~\label{lem:rd}
If $\tuple{l_0, \sigma_0, \pi_0} \Rightarrow ... \Rightarrow \tuple{l_i, \sigma_i, \pi_i}$, then $v \in \dom{\sigma_i}$ implies
   $\exists 0 \leq k < i .  l_k:v := e \in P$, $\tuple{e, \sigma_k, \pi_k} \Rightarrow \sigma_i(v)$, and $l_k \in \pi_i$. 
\end{lemma}
\noindent \textbf{Proof: (induction on $i$)}\\
Base case ($i=0$): If $i=0$, $\dom{\sigma_0} = \emptyset$ so $v \not\in \dom{\sigma_0}$. \\
Induction step (assume for $i$, prove for $i+1$): 
$\tuple{l_i, \sigma_i, \pi_i} \Rightarrow \tuple{l_{i+1}, \sigma_{i+1}, \pi_{i+1}}$. \\
Case analysis on $l_i:c \in P$:
\begin{itemize}
	\item $l_i:c=l_i:w:=e \in P$:
	\begin{itemize}
		\item Known facts:
			\begin{itemize}
                                \item $\sigma_i[w \mapsto n] = \sigma_{i+1}$ where $\tuple{e,\sigma_i, \pi_i} \Rightarrow n$.
				\item $\pi_i[w \mapsto \setof{l_i}] \leq \pi_{i+1}$ by the augmented operational semantics.
			\end{itemize}
                \item Must show: $\forall v \in \dom{\sigma_{i+1}} . \exists 0 \leq k < i+1 .  l_k:v := e \in P$, 
                    $\tuple{e, \sigma_k, \pi_k} \Rightarrow \sigma_{i+1}(v)$, and $l_k \in \pi_{i+1}$.
		 \item For $v = w$, $k = i$:  $v \in \dom{\sigma_{i+1}}$, $l_i : v := e \in P$, $\sigma_{i+1}(v) = n$ where $\tuple{e, \sigma_i, \pi_i} \Rightarrow n$, and $l_i \in \pi_{i+1}$. . 
                 \item For $v \neq w$, if $v \in \dom{\sigma_{i+1}}$, then $v \in \dom{\sigma_i}$, $\sigma_{i+1}(v) = \sigma_i(v)$, $\pi_{i+1}(v) = \pi_i(v)$ and the theorem holds by the induction hypothesis. 
        \end{itemize}

	\item $l_i:c=l_i:\mathbf{if}\; b \; \mathbf{then}\; g \in P$, $l_i:c = l_i:\mathbf{goto}\; g\in P$, $l_i:c = l_i:\mathbf{skip} \in P$, or 
          $l_i:c = l_i:\mathbf{halt} \in P$:
        Then $\pi_{i+1} = \pi_i$ and $\sigma_{i+1} = \sigma_i$ and the theorem holds by the induction hypothesis. 
\end{itemize}
\noindent$\blacksquare$

\comment{
We next prove a lemma related to the relationship between reaching definitions and
constant propagation, specifically that if an assignment
$l:v:= n \in P$ of a constant $n$ to $v$ is recorded as the reaching definition for $v$ in $\pi$ 
(i.e., if $\pi(v) = \setof{l}$), then the state $\sigma$ records $n$ as the value
of $v$:
\begin{lemma}~\label{piConstProp}
	Assume $l$:$v:=n$ $\in P$. Then $\tuple{l_0, \sigma_0, \pi_0} \Rightarrow ... \Rightarrow \tuple{l_i, \sigma_i, \pi_i}$ and $\pi_{i}(v)=\setof{l}$ implies  $\sigma_i(v)=n$
\end{lemma}
\noindent \textbf{Proof: (induction on $i$)}\\
Base case ($i=0$): If $i=0$, $\pi_0(v) = \emptyset$. \\
Induction step (assume for $i$, prove for $i+1$):
$\tuple{l_i, \sigma_i, \pi_i} \rightarrow \tuple{l_{i+1}, \sigma_{i+1}, \pi_{i+1}}$. \\
Case analysis on $l_i:c \in P$:
\begin{itemize}
	\item $l_i:c=l_i:w:=e \in P$:
	\begin{itemize}
		\item Known facts:
			\begin{itemize}
                                \item $\sigma_{i+1} = \sigma_i[w \mapsto m]$ where $\tuple{e,\sigma} \rightarrow m$.
				\item $\pi_{i+1} = \pi_i[w \mapsto \setof{l_i}]$ by the augmented operational semantics.
                                \item $\pi_{i}(v) = \setof{l}$ implies $\sigma_{i}(v) = n$ by induction hypothesis.
			\end{itemize}
		 \item If $l_{i+1} = l$ then $l_i : v := n \in P$, $\pi_{i+1}(v) = \setof{l}$, and $\sigma_{i+1}(v) = n$.
                 \item If $l_{i+1} \neq l$ and $v = w$, then $\pi_{i+1}(v) = \setof{l_i}\neq \setof{l}$. 
                 \item If $l_{i+1} \neq l$ and $v \neq w$, then $\pi_{i+1}(v) = \pi_i(v)$ and $\sigma_{i+1}(v) = \sigma_i(v)$.
                       So by induction hypothesis, $\pi_{i+1}(v) = \setof{l}$ implies $\sigma_{i+1}(v) = n$. 
        \end{itemize}

	\item $l_i:c=l_i:\mathbf{if}\; b \; \mathbf{then}\; g \in P$, $l_i:c = l_i:\mathbf{skip} \in P$, or $l_i:c = l_i:\mathbf{goto}\; g\in P$: \\
        Then $\pi_{i+1} = \pi_i$ and $\sigma_{i+1} = \sigma_i$.  So by induction hypothesis, $\pi_{i+1}(v) = \setof{l}$ implies $\sigma_{i+1}(v) = n$.
\end{itemize}
\noindent$\blacksquare$
}

\subsubsection{Reaching Definitions Analysis}

The program analysis obtains the analysis
results $\beta_{\before{l}}$ and $\beta_{\after{l}}$ 
by formulating and solving, using standard least fixed-point
techniques, the following set of forward dataflow equations:
\[
\begin{array}{rcl}
\beta_{\before{\first{P}}} & = & \lambda v . \emptyset \\
\beta_{\before{l}} & = &  \lor \beta_{\after{g}}, \mbox{ where } l:c \in P \mbox{ and } g \in \pred{l} \\
\beta_{\after{l}} & = &  f(l, \beta_{\before{l}})
\end{array}
\]
\noindent where $f$ is the transfer function for the analysis defined as follows:

\[
f(l, \beta) = 
\begin{cases}
\beta[v\mapsto \setof{l}] & \mbox{if }  l:v:=e \in P \\
\beta & \mbox{otherwise}
\end{cases}
\]

Note that because the analysis is a may analysis (it only
computes definitions that may reach program points), it does not attempt to determine
if that any definition will reach any specific program point --- it only verifies that if a definition
does reach a program point, it will be one of the definitions recorded in the 
corresponding analysis result at that program point. It is, of course, possible to combine the reaching
definitions analysis with the defined variables analysis (Section~\ref{sec:dv}) to obtain
a guarantee that 1) a variable $v$ is always defined at a program point and 2) therefore in all
executions one of the recorded definitions reaches that program point. 

\comment{
We next prove that the analysis computes a conservative overapproximation to the set of reaching
definitions, specifically that the reaching definitions in any execution are a subset
of the reaching definitions computed by the analysis:
\begin{lemma}~\label{pi_leq_sigma_rd}
  If $\tuple{l_0, \sigma_0, \pi_0} \Rightarrow ... \Rightarrow \tuple{l_i, \sigma_i, \pi_i}$,  then $\pi_i \leq \beta_{\before{l_i}}$
\end{lemma}

\noindent\textbf{Proof} (induction on $i$):\\
Base Case $(i=0)$:\\
$\beta_{\before{l_0}} = \lambda v. \emptyset $ and $\pi_0 = \lambda v. \emptyset $, so $\pi_0 \leq \beta_{\before{l_0}}$.\\
Induction Step (assume for $i$, prove for $i+1$):\\
Show $\forall w \in V$, $\pi_{i+1}(w) \subseteq \beta_{\before{l_{i+1}}}(w)$.\\
Case analysis on $l_i:c \in P$:

\begin{itemize}
	\item $l_i:c=l_i:v:=e\in P$:
	\begin{itemize}
		\item Known facts:
			\begin{itemize}
				\item $\pi_i[v \mapsto \{l_i\}]=\pi_{i+1}$ by the augmented operational semantics.
				\item $\beta_{\before{l_{i}}}[v \mapsto \setof{l_i}]  = \beta_{\after{l_i}}$ by the transfer function for $l:v:=e$.
				\item $\beta_{\after{l_i}} \leq \beta_{\before_{l_{i+1}}}$ by the dataflow equations.
				\item $\forall w\in V, \pi_i(w) \subseteq \beta_{\before{i}}(w)$ by the induction hypothesis.
			\end{itemize}
		 	\item Consider any $w \in V$:\\
		 	If $w=v$, then $\pi_{i+1}(w)=\setof{l_i}=\beta_{\after{l_i}}(w) \subseteq \beta_{\before_{l_{i+1}}}(w)$.\\
			If $w \neq v$, then $\pi_{i+1}(w) = \pi_i(w) \subseteq \beta_{\before{l_i}}(w)=\beta_{\after{l_i}}(w) \subseteq \beta_{\before{l_{i+1}}}(w)$.
		\end{itemize}
	\item $l_i:c = l_i:\mathbf{if}\; b \; \mathbf{then}\; g \in P$, $l_i:c = l_i:\mathbf{skip} \in P$, or $l_i:c = l_i:\mathbf{goto}\; g\in P$: \\
	By the induction hypothesis, $\pi_i \leq \beta_{\before{l_i}}$.\\
	By the augmented operational semantics, $\pi_i$ = $\pi_{i+1}$ and $\beta_{\before{l_i}} \leq \beta_{\before{i+1}}$ because $l_i \in \pred{l_{i+1}}$.\\
	Then $\pi_{i+1} \leq \beta_{\before{l_{i+1}}}$.
\end{itemize} 
\noindent$\blacksquare$
}

\begin{theorem}
	\label{thm:rdprogress}
	(Progress): If $\tuple{l,\sigma}\rightarrow\tuple{l',\sigma'}$, then
	$\tuple{l,\sigma, \beta_{\before{l}}}\Rightarrow\tuple{l',\sigma',\beta_{\before{l'}}}$.
\end{theorem}

\noindent{\bf Proof:} 
The standard and augmented program execution rules for $l:c \in P$ have
the same preconditions over and define the same values for $l'$ and $\sigma'$. 
The following case analysis on $l:c \in P$ shows that 
$\beta_{\before{l}}$ and $\beta_{\before{l'}}$ satisfy the history
variable conditions in the augmented operational semantics
so that $\tuple{l, \sigma, \beta_{\before{l}}} \Rightarrow \tuple{l', \sigma', \beta_{\before{l'}}}$:

\begin{itemize}
	\item $l:c = l:v:=e \in P$: 
	\begin{itemize}
		\item Facts from dataflow equations: \\
		$\beta_{\before{l}}[v \mapsto \{l\}] = \beta_{\after{l}}$ (from transfer function $f$ for $l:v:=e \in P$) \\
		$\beta_{\after{l}} \leq \beta_{\before{l'}}$ (because $l \in \pred{l'}$)  
		\item Prove history variable consistency (prove $\beta_{\before{l}}[v \mapsto \{l\}] \leq \beta_{\before{l'}}$): \\
		$\beta_{\before{l}}[v \mapsto \{l\}] = \beta_{\after{l}}$ and $\beta_{\after{l}} \leq \beta_{\before{l'}}$ imply $\beta_{\before{l}}[v \mapsto \{l\}] \leq \beta_{\before{l'}}$.
		\item By the program execution rule for $l:v:=e \in P$, with $l' = \mathbf{next}(l)$, $\sigma' = \sigma[v \mapsto n]$, $\pi = \beta_{\before{l}}$, and $\pi' = \beta_{\before{l'}}, \tuple{l, \sigma, \beta_{\before{l}}} \Rightarrow \tuple{l', \sigma', \beta_{\before{l'}}}$.
	\end{itemize}
	\item $l:c = l:\mathbf{if}\; b \; \mathbf{then}\; g \in P$,  $l:\mathbf{goto}\; g\in P$, $l:\mathbf{skip} \in P$, $l:\mathbf{halt} \in P$:
	\begin{itemize}
		\item Facts from dataflow equations: \\
		$\beta_{\before{l}} = \beta_{\after{l}}$ (from transfer function $f$ for $l:c \in P$) \\
		$\beta_{\after{l}} \leq \beta_{\before{l'}}$ (because $l \in \pred{l'}$)  
		\item Prove history variable consistency (prove $\beta_{\before{l}} \leq \beta_{\before{l'}}$): \\
		$\beta_{\before{l}} = \beta_{\after{l}}$ and $\beta_{\after{l}} \leq \beta_{\before{l'}}$ imply $\beta_{\before{l}} \leq \beta_{\before{l'}}$
		\item By program execution rule for $l:\mathbf{if}\; b \; \mathbf{then}\; g \in P$:
		\begin{itemize}
			\item if $\tuple{l,\sigma,b} \Rightarrow\textbf{true}, l' = g$, $\sigma' = \sigma$, $\pi = \beta_{\before{l}}$ and $\pi' = \beta_{\before{l'}}, \tuple{l, \sigma, \beta_{\before{l}}} \Rightarrow \tuple{l', \sigma', \beta_{\before{l'}}}$.
			\item if $\tuple{l,\sigma,b} \Rightarrow\textbf{false}, l' = \mathbf{next}(l)$, $\sigma' = \sigma$, $\pi = \beta_{\before{l}}$ and $\pi' = \beta_{\before{l'}}, \tuple{l, \sigma, \beta_{\before{l}}} \Rightarrow \tuple{l', \sigma', \beta_{\before{l'}}}$.
		\end{itemize}
		\item By program execution rule for $l:\textbf{goto}\;g \in P$, with  $l' = g$, $\sigma' = \sigma$, $\pi = \beta_{\before{l}}$, and $\pi' = \beta_{\before{l'}}, \tuple{l, \sigma, \beta_{\before{l}}} \Rightarrow \tuple{l', \sigma', \beta_{\before{l'}}}$
		\item By program execution rule for $l:\mathbf{skip} \in P$ or $l:\mathbf{halt} \in P$
                with  $l' = \mathbf{next}(l)$, $\sigma' = \sigma$, $\pi = \beta_{\before{l}}$, and $\pi' = \beta_{\before{l'}}$:
		$\tuple{l, \sigma, \beta_{\before{l}}} \Rightarrow \tuple{l', \sigma', \beta_{\before{l'}}}$
	\end{itemize}
\end{itemize}
\noindent$\blacksquare$

\subsubsection{Reaching Definitions Correctness Theorem}

We next use the Progress theorem (Theorem~\ref{thm:rdprogress}) to illustrate the application
of reaching definitions to constant propagation. The theorem states that if all of the
definitions of a variable $v$ that reach a given program point are from assignments 
of $v$ to the same constant $n$, then in any execution of the program at that point,
if the value of $v$ is defined, then the value of $v$ is $n$:
\begin{theorem}
If $\tuple{l_0, \sigma_0} \rightarrow \ldots \rightarrow \tuple{l_i, \sigma_i}$, $v\in \dom{\sigma_i}$, and
$\forall g \in \beta_{\before{l_i}}(v). g:v:=n \in P$, then $\sigma_i(v) = n$.
\end{theorem}
\noindent \textbf{Proof:}
If $\tuple{l_0, \sigma_0} \rightarrow \ldots \rightarrow \tuple{l_i, \sigma_i}$, then by Progress (Theorem~\ref{thm:rdprogress}), 
$\tuple{l_0, \sigma_0, \beta_{\before{l_0}}} \Rightarrow \ldots \Rightarrow \tuple{l_i, \sigma_i, \beta_{\before{l_i}}}$
where $\beta_{\before{l_0}} = \pi_0$. 
By Lemma~\ref{lem:rd}, $v \in \dom{\sigma_i}$ implies $\exists 0 \leq k < i.l_k:v := e \in P$, $l_k \in \beta_{\before{l_i}}$, 
and $\tuple{e,\sigma_k} \rightarrow \sigma_i(v)$. 
Consider $l_k$. Because $l_k \in \beta_{\before{l_i}}$, $l_k : v := n \in P$, $\tuple{e,\sigma_k} \rightarrow n$, and $\sigma_i(v) = n$. 
\noindent$\blacksquare$

The defined variables analysis (Section~\ref{sec:dv}) is designed to determine
if a variable $v$ is always defined at a given program point. If so, the value
of $v$ at that point is always given by one of the definitions identified
by the reaching definitions analysis. 

	\section{Related Work}
\label{sec:related}

Simulation relations, and techniques for proving that simulation relations exist, have
been extensively explored in the context of establishing simulation relations
between state machines~\cite{LynchV95,LynchDistributedAlgorithms}. The developed theory includes a range
of proof techniques and mechanisms, including forward and backward proof techniques
with refinement mappings, abstraction functions, and abstraction relations. 
Prophecy variables were initially developed for the purpose of proving
that implementations satisfy specifications via refinement
mappings with forward simulations, specifically in the case when the 
specification makes a choice before the implementation~\cite{AbadiL91}.
The addition of prophecy variables to the framework of refinement mappings
with history variables and forward simulation proofs enabled a completeness
result for the ability to prove trace inclusions of implementations within 
specifications~\cite{AbadiL91}. It is, of course, known that backward
simulation is an alternative to forward simulation with prophecy 
variables~\cite{LynchV95}. In general, there are a number of alternatives 
when choosing a formal framework for proving simulation properties, with the
appropriate framework depending on pragmatic issues
such as the convenience and conceptual difficulty of working with the 
concepts in the framework. In general, approaches that reason forward
in time seem to be more attractive and intuitive than approaches that
reason backward against time, as can be seen, for example, in pedagogical
presentations of dataflow analyses, which invariably present forward
analyses first, then backward analyses second as a kind of dual of forward 
analyses~\cite{appel2004modern,muchnick1997advanced,DragonBook,cooper2011engineering,kennedy2001optimizing,Aldrich2019Correctness,Aldrich2019Examples}.

Many of the concepts that appear in simulation relation proofs for state
machines also appear in the program verification, dataflow analysis, 
and abstract interpretation literature.
For example, history variables were first introduced in the program
verification literature~\cite{OwickiG76}, abstraction functions,
originally introduced in the program verification literature~\cite{Hoare72}, 
can be seen as a form of refinement mappings, 
and program analyses can be seen as establishing a simulation relation between an
abstract interpretation of the program (which plays the role of the
specification) and concrete executions of the program (which play the
role of the implementation)~\cite{CousotC77,CousotC92}. It is also known that, in this context,
backward or reverse simulation relations can be used to establish
the correspondence between backward analyses (which extract
information about the future execution) and program 
executions~\cite{CousotC92,SchmidtS98}.

In this paper we introduce prophecy variables to enable forward reasoning about 
program analysis properties that involve the future execution of the program. 
To the best of our knowledge, we are the first to introduce prophecy variables for
this purpose (here we contrast with the recent use of prophecy variables for
program verification~\cite{JungLPRTDJ20,Vafeiadis08,ZhangFFSL12} as
well as the traditional use of prophecy variables for proving forward simulation
relations between state machines~\cite{AbadiL91}). In this context prophecy
variables enable a unified treatment of forward and backward dataflow analyses
and support forward reasoning to establish correctness properties that involve
backward analysis results (Theorems~\ref{thm:lvcharacterize}, \ref{thm:vbeca}, \ref{thm:vbech}).

We also exploit aspects of the program analysis context to specialize the
more general state machine simulation relation framework to the program
analysis context.  The result is a simpler and more tractable framework 
as appropriate in this context:
\begin{itemize}
\item Drawing the prophecy and history variables $\pi$ and the
analysis results $\beta$ from the same lattice eliminates the need to 
work with an explicit abstraction function or refinement mapping $\alpha$ 
to establish a connection between the analysis and program executions.
The resulting direct connection between the analysis 
and the execution eliminates the abstraction function/refinement mapping
from proofs that connect the analysis with the execution and from any
subsequent correctness proofs involving the analysis results. 

\item Instead of using a refinement mapping or abstraction function
to establish a one-way simulation relation between 
a specification and an implementation or between concrete and abstract
executions of the program, in our approach correct analysis
results establish a two-way bisimulation between the standard semantics
and the augmented semantics over the analysis results $\beta$. 

\item Augmenting the standard operational semantics with prophecy or
history variables $\pi$ eliminates the need to work with traditional instrumented
or collecting semantics --- the prophecy or history variable updates
(which typically parallel the updates to the standard program state $\sigma$)
directly extract this information as the program executes. It is possible
to see the prophecy and history variable mechanism in this context as
replacing the combination of a traditional instrumented/collecting semantics plus
an abstraction function with a single unified mechanism. 

\end{itemize}

Prophecy variables have recently been applied for program verification in 
a Hoare program logic based on separation logic~\cite{JungLPRTDJ20}.
Our purpose is different, specifically to use prophecy variables to 
enable forward reasoning in the context of backward dataflow analysis algorithms.
Instead of using prophecy variables to enable forward reasoning about
properties of complex parallel algorithms and data structures,
we use prophecy variables to prove properties of algorithms that analyze sequential 
programs (as well as properties of the analysis results that they produce). 

Cobalt enables compiler developers to specify a range of dataflow optimizations
(such as constant propagation and partial dead assignment elimination)~\cite{LernerMC03}.
Each optimization is specified by a transformation pattern whose guard specifies
a condition over sequences of actions in paths in the program representation that must hold for the transformation
to be legal. Cobalt has separate constructs for specifying forward and backward 
optimizations --- forward guards reason about forward properties, backward guards
reason about backward properties. 

Rhodium implements soundess proofs for dataflow analyses~\cite{LernerMRC05}. It largely automates
a standard dataflow analysis setup, with de facto abstraction functions (expressed as 
predicates over concrete program states) establishing the connection
between concrete program states and dataflow facts and state extensions
(a form of instrumented semantics) to support analyses that extract information 
about the past execution of the program not present in standard concrete program
states. Like Cobalt, Rhodium has separate support for forward and backward analyses;
subsequent work on automatically inferring correct propagation rules supports
only forward rules~\cite{ScherpelzLC07}.

We note that dataflow analysis and abstract interpretation are large
fields with a long history of technical development. In this work we
aspire only to rework the treatment of some of the basic concepts in the field. We note that integrating
backward and forward information via alternating backward and
forward analyses is a known technique~\cite{CousotC99},
including transformations of analyzed systems of Horn clauses
to effectively convert combined backward and forward 
Horn clause analysis problems into forward analysis problems~\cite{KafleG15,BakhirkinM17}.
It remains to be seen what, if any, role prophecy variables may usefully play
in combining these kinds of backward and forward analysis problems. 

Researchers have also formulated dataflow correctness properties via
temporal logic~\cite{Steffen91,SchmidtS98}, which can be seen as specifying properties about
paths that connect relevant program actions, such as writing
or reading a variable, in the representation of the program. 
The approach can therefore eliminate the need for an 
instrumented operational semantics that explicitly carries
information about the past execution through the program
representation. In our approach this kind of information 
(when required) is stored explicitly in
prophecy and history variables, propagated locally, and
updated by the augmented operational semantics. 

The CompCert verified compiler contains an implementation of a 
generally standard dataflow analysis framework for supporting
traditional compiler optimizations such as constant propagation
and common subexpression elimination~\cite{BertotGL04}.
The formulation includes lattices of dataflow facts, 
abstraction functions for mapping register values to lattice values, 
and a forward and backward implementation of Kildall's
fixed point algorithm for solving dataflow equations. 
Example dataflow domains record when registers contain constant
values (for constant propagation) or the expressions for register
values (for common subexpression elimination). 

	\section{Conclusion}
\label{sec:conclusion}

Dataflow analysis has been the focus of intensive research for decades. Despite this focus,
and despite conceptual similarities between many problems that arise 
in program analysis and state machine refinement proofs, prophecy variables
(originally developed to support forward state machine simulation relation
proofs) have seen little to no application to program analysis problems.
By showing how to use prophecy variables to enable forward reasoning for
backward dataflow analyses, as well as developing a streamlined treatment
of both backward and forward dataflow analyses based on prophecy and history
variables, we hope to promote the use of these mechanisms as appropriate
to productively revisit basic concepts in the field and obtain a more unified and effective approach
to a range of program analysis problems.

	\bibliography{paper.bib}

\end{document}